\documentclass[12pt,a4paper]{article}
\pdfoutput=1
\usepackage[numbers,compress,square]{natbib}
\usepackage[utf8]{inputenc}
\usepackage[T1]{fontenc}
\usepackage{style}
\usepackage[english]{babel}
\usepackage{amsmath}
\usepackage{graphicx} 
\usepackage{verbatim} 
\usepackage{mathtools}
\usepackage{amssymb}
\usepackage{url}
\usepackage{footnote}
\usepackage{booktabs}
\usepackage{bm}
\usepackage{enumitem}
\usepackage{caption}
\usepackage{subcaption}

\DeclarePairedDelimiter{\abs}{\lvert}{\rvert}

\renewcommand{\Im}{\mathop{\rm Im}\nolimits}

\def\half{\frac{1}{2}}

\def\d{\partial}
\newcommand{\bseq}{\begin{subequations}}
\renewcommand{\comment}[1]{}
\newcommand{\eseq}{\end{subequations}}
\newcommand{\be}{\begin{equation}}
\newcommand{\ee}{\end{equation}}
\newcommand{\beqa}{\begin{eqnarray}}
\newcommand{\eeqa}{\end{eqnarray}}

\newcommand\p{{\bf p}}

\newcommand\x{{\bf x}}
\renewcommand\k{{\bf k}}
\newcommand\q{{\bf q}}

\renewcommand{\ln}{\mathop{\rm ln}\nolimits}

\renewcommand{\k}{{\bf k}}

\newcommand{\astfootnote}[1]{%
\let\oldthefootnote=\thefootnote%
\setcounter{footnote}{0}%
\renewcommand{\thefootnote}{\fnsymbol{footnote}}%
\footnote{#1}%
\let\thefootnote=\oldthefootnote%
}

\newcommand{\hp}{\hphantom}
\makeatletter
\newlength{\apb@width}
\newcommand{\autoparbox}[2][c]{\settowidth{\apb@width}{#2}\parbox[#1]{\apb@width}{#2}}

\makeatother
\newcommand\eq[1]{Eq.~\eqref{eq:#1}}

\newcommand{\eqsII}[2]{Eqs.~\eqref{eq:#1}, \eqref{eq:#2}}

\newcommand{\iu}{i}
\newcommand{\eu}{e}
\newcommand{\dif}{d}
\newcommand\del{\partial}

\renewcommand{\vec}{\bm} 
\newcommand\vers[1]{\hat{\vec{#1}}}

\def\Im{\rm Im}
\def\mpl{M_{\rm P}}

\newcommand\qvec{{\bf q}}
\newcommand\kvec{{\bf k}}
\newcommand\xvec{{\bf x}}

\newcommand{\andd}{\ , \quad \text{and}  \quad}

\newcommand{\secref}[1]{Sec.~\ref{#1}}

\newcommand{\figref}[1]{Fig.~\ref{#1}}

\newcommand{\eqn}[1]{Eq.~(\ref{#1})}

\newcommand{\unitsk}{\, h \, { \rm Mpc^{-1}}}
\newcommand{\lcdm}{$\Lambda$CDM }

\newcommand{\omegam}{\Omega_{\rm m,0}}  

\def\MM{M_{*}}
\newcommand{\R}{R}
\newcommand\alphaB{\alpha_{\text{B}}}
\newcommand\alphaM{\alpha_{\text{M}}}

\newcommand\alphaT{\alpha_{\text{T}}}
\newcommand\alphaH{\alpha_{\text{H}}}
\newcommand\bone{{\alpha}_{\rm V1}}
\newcommand\btwo{{\alpha}_{\rm V2}}
\newcommand\bthree{{\alpha}_{\rm V3}}
\newcommand{\MP}{M_{\rm P}}
\newcommand{\cH}{\mathcal{H}}
\newcommand{\knl}{k_{\rm NL}}
\newcommand{\epsilonosc}{\epsilon_{\rm osc}}

\relax

\title{Snowmass White Paper: Effective Field Theories in Cosmology}

\author[a]{Giovanni Cabass,}
\author[a]{Mikhail M. Ivanov,}
\author[b]{Matthew Lewandowski,}
\author[c]{Mehrdad Mirbabayi,}
\author[d]{and Marko Simonovi\'c}

\affiliation[a]{School of Natural Sciences, Institute for Advanced Study,\\1 Einstein Drive, Princeton, NJ 08540, USA}
\affiliation[b]{Department of Physics and Astronomy,
Northwestern University, Evanston, IL 60208}%
\affiliation[c]{Abdus Salam International Centre for Theoretical Physics,\\Strada Costiera 11, 34151, Trieste, Italy}%
\affiliation[d]{Theoretical Physics Department, CERN,\\1 Esplanade des Particules, Geneva 23, CH-1211, Switzerland}

\abstract{Small fluctuations around homogeneous and isotropic expanding backgrounds are the main object of study in cosmology. Their origin and evolution is sensitive to the physical processes that happen during inflation and in the late Universe. As such, they hold the key to answering many of the major open questions in cosmology. Given a large separation of relevant scales in many examples of interest, the most natural description of these fluctuations is formulated in terms of effective field theories. This was the main avenue for many of the important modern developments in theoretical cosmology, which provided a unifying framework for a plethora of cosmological models and made a clear connection between the fundamental cosmological parameters and observables. In this review we summarize these results in the context of effective field theories of inflation, large-scale structure, and dark energy.

\vspace{2.5cm}

\begin{center}
  \rule[-0.2in]{\hsize}{0.01in}\\
  \rule{\hsize}{0.01in}\\
  \vskip 0.1in
  Submitted to the Proceedings of the US Community Study\\ 
  on the Future of Particle Physics (Snowmass 2021)\\
  \rule{\hsize}{0.01in}\\
  \rule[+0.2in]{\hsize}{0.01in}\\[-2em]
\end{center}

}

\begin{document}

\maketitle
\flushbottom

\section{Executive Summary}

\noindent During the last decade we witnessed a large progress in application of effective field theory (EFT) techniques in cosmology. The main object of study of these EFTs are small cosmological perturbations, their evolution and interactions on scales relevant for cosmology. Examples of such perturbations include quantum-mechanical fluctuations of the inflaton which provide the seeds for small density fluctuations in the late Universe, fluctuations in temperature and polarization of the cosmic microwave background (CMB), fluctuations in the number density of galaxies in the cosmic web at low redshifts which forms the large-scale structure (LSS), and fluctuations in some hypothetical medium which drives the current accelerated expansion of the Universe (i.e.~dark energy (DE)). Just like in any other generic EFT, in all these examples the action for the relevant degrees of freedom can be expanded in powers of small fluctuations and derivatives and at every order its form is fixed by the symmetries up to a finite number of free coefficients. This description is valid up to a cutoff, where the details of the UV physics become relevant. However, below such cutoff (on large enough scales) the EFT predictions are universal. Different cosmological models or different UV completions differ only by different values of the EFT parameters.

The EFT approach has two major advantages. First, the EFT provides a unified description of many different particular models, keeping only their essential properties and identifying long-wavelength degrees of freedom relevant for cosmology. At the same time, it also allows for a clear separation of the theory of small fluctuations around homogenous background from the evolution of the background itself. Second, EFTs in cosmology are weakly coupled theories, hence they can be used to make perturbative predictions for all relevant observables throughout the entire history of the Universe, from the Bunch-Davies vacuum in inflation to observed galaxy distribution at present times. More precisely, at each order in perturbation theory and derivative expansion, one can calculate a finite number of ``shapes,’’ i.e.~momenta dependence, of observable $n$-point correlation functions whose amplitudes are proportional to the free EFT coefficients. Importantly, such calculations can be always perturbatively improved to match the precision required by the statistical errors of a given experiment. These shapes can be then used for comparison to the data. In this way the EFT approach not only played an important conceptual role of simplifying theoretical calculations and unifying different cosmological models, but also made a large impact on observational cosmology, inspiring templates used in the data analysis and providing a way for robust measurements of cosmological parameters, allowing for an easy marginalization over the unknown UV physics.

Historically, the EFT methods in cosmology were first applied to inflation~\cite{Creminelli:2006xe,Cheung:2007st}. We will mainly focus on the simplest incarnation, the effective field theory of single-field inflation (EFTI), where inflation is driven by a single medium whose quantum fluctuations produce the observable overdensities in the Universe. At the time when the EFTI appeared, it unified a rapidly growing number of inflationary models and provided a simple Lagrangian for inflaton fluctuations. Using symmetry arguments, the EFTI provided a clear connection between possibly small speed of sound of inflaton perturbations and large primordial non-Gaussianities (PNG) of equilateral and orthogonal shapes~\cite{Babich:2004gb,Senatore:2009gt}. This result paved the road for observational constraints on speed of propagation of inflaton fluctuations, which in turn can tell us a lot about the  physics of inflation~\cite{Creminelli:2003iq,Baumann:2011su}. Furthermore, the existence of new shapes of potentially large PNG in single-field models (the local shape was known to be absent in single-field inflation~\cite{Maldacena:2002vr,Creminelli:2004yq}), gave an additional boost to the phenomenology of PNG. The EFTI was since then generalized to include multi-field models and other extensions of the basic single-field inflation, playing the role of a common language in the field of primordial cosmology. Many of the recent developments, such as the study of imprints of massive and higher spin particles on cosmological correlation functions or cosmological bootstrap, are motivated by what we have learned from the EFTI.

Another application of the EFT in cosmology, which is becoming increasingly more important in recent years, is to  galaxy clustering and the LSS of the Universe. With the ever growing data sets where the spectroscopic galaxy samples increase by a factor of 10 every decade, ongoing and upcoming galaxy surveys have the potential to become one of the leading probes of cosmology, reaching and even surpassing the precision of the CMB observations. The effective field theory of large-scale structure (EFT of LSS)~\cite{Baumann:2010tm,Carrasco:2012cv,Perko:2016puo} is perfectly placed to face the challenge of interpreting this large amount of data. Being an EFT of fluctuations of the number density of galaxies (or other tracers of matter), it allows for a systematic description of galaxy clustering on large scales, regardless of complicated galaxy formation which strongly depends on details of poorly understood baryonic physics. The cutoff of the theory is given by the scale where gravitational nonlinearities and feedback from astrophysical processes become large and it is typically of the order of a few megaparsecs. Dynamics on larger scales is driven only by gravitational interactions and all UV physics can be captured in effective contributions to the equations of motion that are organized as an expansion in the number of fields and derivatives. Such description, which radically separates galaxy formation physics from the long-wavelength dynamics of fluctuations in the number density of galaxies, proved to be extremely useful in practice. Even though a lot of work still remains to be done, already the consistent leading EFT calculations, such as the one-loop power spectrum and tree-level bispectrum, led to important advancements in the program of obtaining cosmological information from the LSS galaxy surveys. Those include the first inference of all fundamental cosmological parameters from the galaxy power spectrum~\cite{Ivanov:2019pdj,DAmico:2019fhj,Chen:2021wdi} and the first constraints on primordial non-Gaussianity from the galaxy bispectrum~\cite{Cabass:2022wjy,DAmico:2022gki}, two major milestones that were elusive for a long time in the past. Further theoretical improvements and theory-inspired novel data analysis techniques can lead to further progress and this remains a very active area of research.

Finally, EFT methods were also applied in the context of DE. Assuming that DE is a medium that drives the current accelerated expansion of the Universe, but that can also fluctuate, one can formulate the effective field theory of dark energy (EFT of DE)~\cite{Gubitosi:2012hu} in a way similar to the EFTI. One important difference is that the couplings of the dark energy field to the matter fields have to be carefully taken into account. As in the other two examples, without the need to refer to any UV physics, one can produce a consistent EFT description that encapsulates all possible phenomenology of DE beyond the cosmological constant. This is very important, since the EFT formulation allows us to consistently parametrize any deviation from the $\Lambda$CDM cosmological model which are compatible with all symmetries and general principles of physics. This in turn is a crucial input for exploring and constraining dark energy properties through observations of galaxy clustering on large scales, one of the key science goals for many galaxy surveys in this decade. In parallel, the EFT of DE is formulated in a way which allows for a straightforward connection of its predictions to relevant astrophysical observations, such as mergers of black holes  and neutron stars. This led to a burst of activity where the detection of gravitational waves was used to put constraints on the EFT parameters~\cite{Creminelli:2017sry, Sakstein:2017xjx, Ezquiaga:2017ekz}. Many other interesting theoretical questions, such as constraints on the EFT parameters from positivity bounds, will remain a playground for fruitful collaborations of high-energy physicists and cosmologists in the years to come.

In conclusion, EFT methods play the central role in theories of cosmological perturbations and as such they are the key in connecting theory and observations and pivotal for answering all the biggest open questions in cosmology. These include the physics of inflation, properties of dark matter and dark energy, and possible discovery of new, additional energy components in our Universe and new physical processes related to them. In this review we summarize the current status of EFTs in cosmology, focusing on three influential examples: effective field theory of inflation, effective field theory of large-scale structure, and effective field theory of dark energy. We present the most important results, connection to cosmological observables, some open problems and directions for future research as well as connections to neighbouring fields of high-energy physics and astrophysics.

\section{Effective Field Theory of Inflation}
\label{sec:efti}


\noindent The energy available to processes during inflation could have been as high as $10^{14}\,{\rm GeV}$, 
far beyond what can be achieved in particle accelerators. Interactions between 
the degrees of freedom active during inflation leave their imprints in the statistics of cosmological perturbations, 
like anisotropies in the CMB temperature and inhomogeneities in the distribution of galaxies, 
therefore offering a privileged view on these energy scales (for prospects of constraining inflation using these observations, see also the snowmass white paper on inflation~\cite{Snowmass2021:Inflation} and references therein).

What are the light degrees of freedom during inflation? We know that at least one scalar 
degree of freedom must have been present. The epoch of accelerated expansion eventually ends, 
so there must have been a ``clock'' that tracks the transition to a decelerated Universe. 
The fluctuations of this clock, together with the fluctuations of the metric, are the degrees of freedom that 
are guaranteed to be active during inflation. The simplest effective field theory of inflation is one for this degree of freedom: it provides a unified description of all inflationary models where inflation is driven by a single clock. Additional light degrees of freedom are included in the EFT following the same general principles.

\subsection{Unitary-gauge action} 
\label{subsec:unitary_gauge_action}

\noindent How can we write an action that encompasses all the single-clock inflationary models? 
Refs.~\cite{Creminelli:2006xe,Cheung:2007st} showed how to achieve this. By using the 
freedom of changing coordinate system, the fluctuations of the clock can be absorbed by the metric. In models where the clock is a scalar field $\phi$, i.e.~the \emph{inflaton}, we can write 
$\phi(t,\xvec) = \phi_0(t) + \delta\phi(t,\xvec)$: the new coordinate system corresponds to 
setting $\delta\phi = 0$. 

In this ``unitary gauge,'' 
the graviton has three degrees of freedom: the scalar mode and two tensor helicities. Writing down the action is now simply a matter of 
finding all operators that are invariant under time-dependent spatial diffeomorphisms, 
since time diffeomorphisms have been fixed. 

\begin{figure}[t]
\begin{minipage}[b]{0.60\linewidth}
\centering
\includegraphics[width=0.9\textwidth]{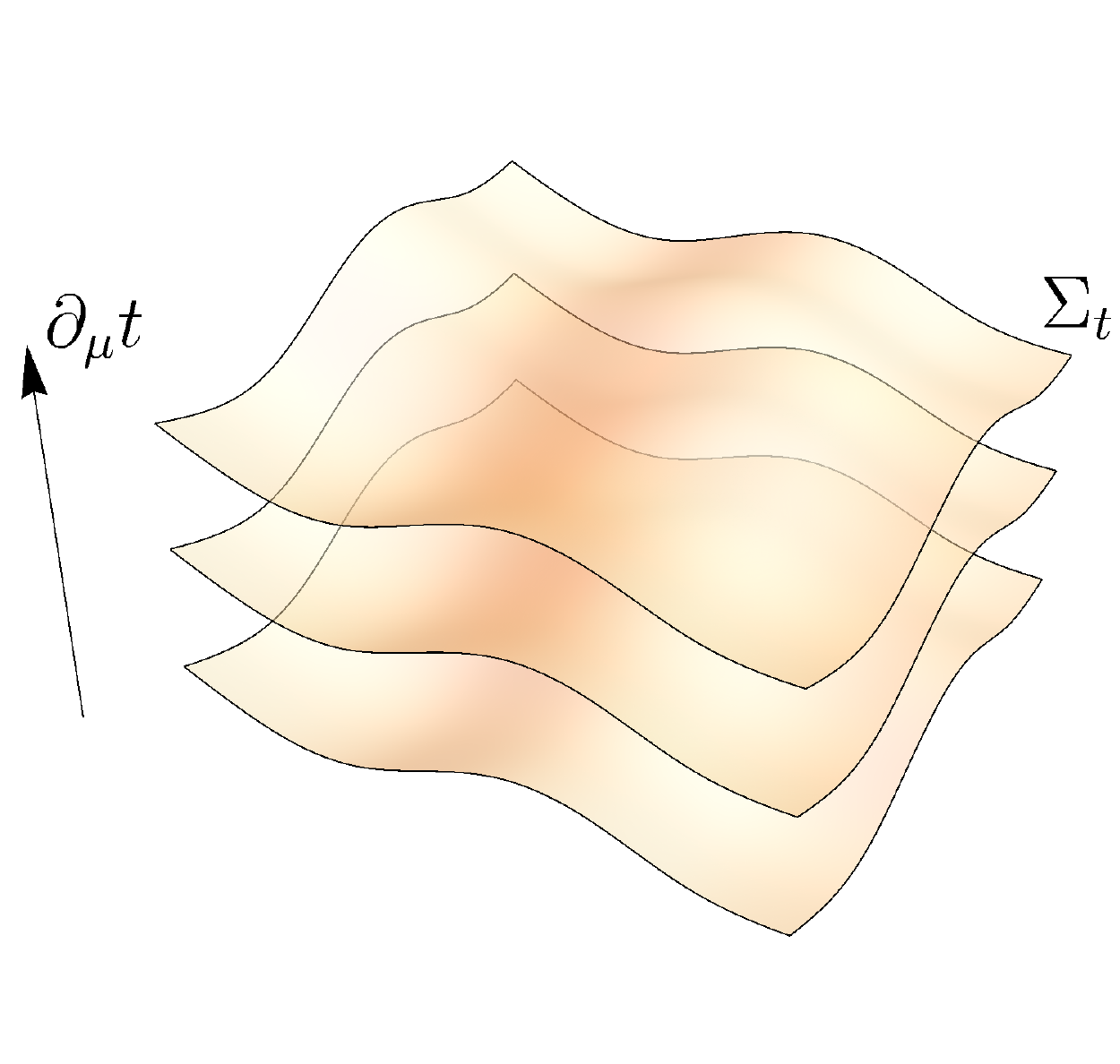}
\par\vspace{0pt}
\end{minipage}
\begin{minipage}[b]{0.30\linewidth}
\centering
\begin{tabular}{cc}
\toprule
$N$ & $\sqrt{-1/g^{00}}$ \\[2ex]
$n_\mu$ & $-N\delta^0_\mu$ \\[2ex]
$h_{\mu\nu}$ & $g_{\mu\nu} + n_\mu n_\nu$ \\[2ex]
$K_{\mu\nu}$ & $h_\mu^{\hp{\mu}\rho}\nabla_\rho n_\nu$ \\ 
\bottomrule
\end{tabular}
\par\vspace{40pt}
\end{minipage}
\caption{\footnotesize $3+1$ decomposition of spacetime, with the expressions for some of the associated geometric objects in terms of the metric: 
the lapse function ($N$), the normal unit vector to constant-time hypersurfaces ($n_\mu$), the projector 
on these hypersurfaces ($h_{\mu\nu}$), and the extrinsic curvature ($K_{\mu\nu}$).} 
\label{fig:3+1}
\end{figure}

The clock defines a preferred foliation of spacetime: the $3+1$ decomposition is 
therefore well-suited to find all operators that are invariant under spatial diffeomorphisms, i.e.~changes of coordinates 
on the hypersurfaces of constant time. This is summarized in Fig.~\ref{fig:3+1} and the accompanying table.

The second step is as follows. We are interested in constructing an effective field theory for cosmological perturbations, 
i.e.~fluctuations around a Friedmann–Lema{\^i}tre\\–Robertson–Walker (FLRW) spacetime.\footnote{We focus on the case of zero spatial 
curvature. See Appendix B of \cite{Cheung:2007st} for how to include it.} 
This is a highly symmetric spacetime, whose line element is written as 
\begin{equation}
\label{eq:efti-1}
\dif s^2 = {-\dif t}^2 + a^2(t)\delta_{ij}\dif x^i\dif x^j\,\,. 
\end{equation} 
The Hubble rate is defined as 
\begin{equation}
\label{eq:efti-2}
H=\frac{\dot{a}}{a}\,\,.
\end{equation} 
$H$ is constant for a de Sitter metric, and the background stress tensor is a cosmological constant. More generally, the background stress tensor can be build from only two operators 
in this $3+1$ decomposition of spacetime: a free function of time $\Lambda(t)$ and the operator $c(t) g^{00}$, 
where $c(t)$ is also a free function. The most general action can then be written as
\begin{equation}
\label{eq:efti-3} 
\begin{split} 
S &= \int\dif^4x\,\sqrt{-g}\,\bigg[\frac{\mpl^2}{2}R - c(t)g^{00} - \Lambda(t) + \frac{M_2^4(t)}{2}\,(g^{00}+1)^2 + \frac{M_3^4(t)}{6}\,(g^{00}+1)^3 \\ 
&\hp{= \int\dif^4x\,\sqrt{-g}\,\bigg[}\,\, - \frac{\bar{M}^3_1(t)}{2}\,(g^{00}+1)\,\delta\!K^\mu_{\hp{\mu}\mu} 
- \frac{\bar{M}^2_2(t)}{2}\,(\delta\!K^\mu_{\hp{\mu}\mu})^2 + \cdots\bigg]\,\,, 
\end{split} 
\end{equation} 
where $\delta\!K_{\mu\nu} = K_{\mu\nu} - H(t)h_{\mu\nu}$. All operators beyond 
the first three have vanishing derivative with respect to $g^{\mu\nu}$ on a FLRW metric. 
As a consequence, $c(t)$ and $\Lambda(t)$ are fixed in terms of the expansion history: 
\begin{subequations}
\label{eq:efti-4}
\begin{align}
c(t) &= {-\mpl^2\dot{H}}\,\,, \label{eq:efti-4-1} \\
\Lambda(t) &= \mpl^2(3H^2+\dot{H})\,\,. \label{eq:efti-4-2}
\end{align}
\end{subequations} 
The difference between different inflationary models is then encoded in the remaining operators, which 
from now on we will call \emph{EFT operators}. 

The organizing principle in \eq{efti-3} is the expansion in perturbations and derivatives, central to all effective field theories. We see that 
the EFT operators are organized by the number of derivatives acting on the unitary-gauge metric and by the order in 
perturbations around an FLRW metric to which they start. 
We will discuss the EFT expansion and the relevant cutoff scales in more detail once we reintroduce the scalar degree of 
freedom via the Stueckelberg trick in Section~\ref{subsec:stueckelberg_trick}. 

The terms ``$\cdots$'' we have not explicitly written in the action of \eq{efti-3} are built not only from $g^{00}$ 
and the extrinsic curvature. Besides these and many other time-diffeomorphisms-breaking operators (see e.g.~Ref.~\cite{Bordin:2017hal} for a 
comprehensive study), 
we also have covariant operators built from the four-dimensional Riemann tensor: these capture corrections to General Relativity.

\subsection{Different models in the EFTI language} 
\label{subsec:models}

The simplest models are those where the clock is the inflaton $\phi$ with minimal kinetic term and potential $V(\phi)$. In the unitary gauge $c(t) = \dot{\phi}^2_0(t)/2$ and $\Lambda(t) = V(\phi_0(t))$, while all the other terms in \eq{efti-3} are set to zero. 
This is the formulation of \emph{slow-roll inflation} in the EFTI \cite{Maldacena:2002vr,Cheung:2007st}.  

Models where there is at most one derivative acting on $\phi$, i.e. 
\begin{equation}
\label{eq:efti-5}
\text{$S_\phi = \int\dif^4x\,\sqrt{-g}\,P(X,\phi)$ \quad with \quad $X = g^{\mu\nu}\del_\mu\phi\del_\nu\phi\,\,,$} 
\end{equation} 
have 
\begin{equation}
\label{eq:efti-6}
M^4_n(t) = \dot{\phi}^{2n}_0(t)\,\frac{\partial^nP}{\partial X^n}\bigg|_{\phi = \phi_0(t)}\,\,, \qquad \bar M_n =0\,\,. 
\end{equation} 
This is \emph{K-inflation}  \cite{Armendariz-Picon:1999hyi,Armendariz-Picon:2000nqq, ArmendarizPicon:1999rj,Garriga:1999vw,Chen:2006nt}. A particular example of $P(X,\phi)$ theory is 
DBI inflation \cite{Alishahiha:2004eh}. 
There the inflaton is the position of a probe brane 
in $5$-dimensional spacetime and its action is constructed from the induced metric on this brane. 
Examples of theories that are described by operators involving $\delta K_{\mu\nu}$ are 
the Ghost Condensate \cite{Arkani-Hamed:2003pdi,Creminelli:2006xe,Cheung:2007st}, Galileons 
\cite{Nicolis:2008in,Burrage:2010cu} and generalizations of DBI Inflation \cite{deRham:2010eu}. 

\subsection{Slow-roll solution and approximate time-translation symmetry} 
\label{subsec:more_unitary_gauge_action}

The coefficients in the unitary gauge action can explicitly depend on time. However, the first two coefficients, $c(t)$ and $\Lambda(t)$, have a mild dependence if the background solution satisfies the slow-roll conditions $\varepsilon\ll 1$, $|\eta|\ll 1$, where
\begin{equation}
\label{eq:efti-7}
\varepsilon \equiv {-\frac{\dot{H}}{H^2}}\,\,,\qquad 
\eta \equiv \frac{\dot\varepsilon}{H\varepsilon}.
\end{equation} 
It is natural to assume that the same holds for all the other coefficients. Namely, to impose an approximate time-translation symmetry, which in slow-roll models follows from the approximate shift symmetry of the inflaton $\phi$. 

{An exception to this rule comes from models where instead of a softly broken continuous shift symmetry, one has a \emph{discrete} one 
\cite{Flauger:2009ab,Flauger:2010ja}. Ref.~\cite{Behbahani:2011it} explored this in the context of the EFTI: at the level of the unitary-gauge action the approximate discrete shift symmetry corresponds to an expansion history $H(t)$, and other time-dependent coefficients, that are a superposition like} 
\begin{equation} 
\label{eq:efti-resonant}
H(t) = H_{\rm sr}(t) + H_{\rm osc}(t)\sin\omega t\,\,,
\end{equation}
where $H_{\rm sr}\gg H_{\rm osc}$ have a slow time dependence of order $\varepsilon$. See Refs.~\cite{Planck:2018jri,Akrami:2019izv} for CMB constraints on oscillating features predicted by these models. 

Another situation where time-translation symmetry is not a good approximation is the ``ultra-slow-roll'' phase \cite{Namjoo:2012aa}. 
See \cite{Finelli:2018upr} and references therein for further discussion.

\subsection{Observables and primordial non-Gaussianity}
\label{subsec:observables}

\begin{figure}[t]
\centering
\includegraphics[width=0.9\textwidth]{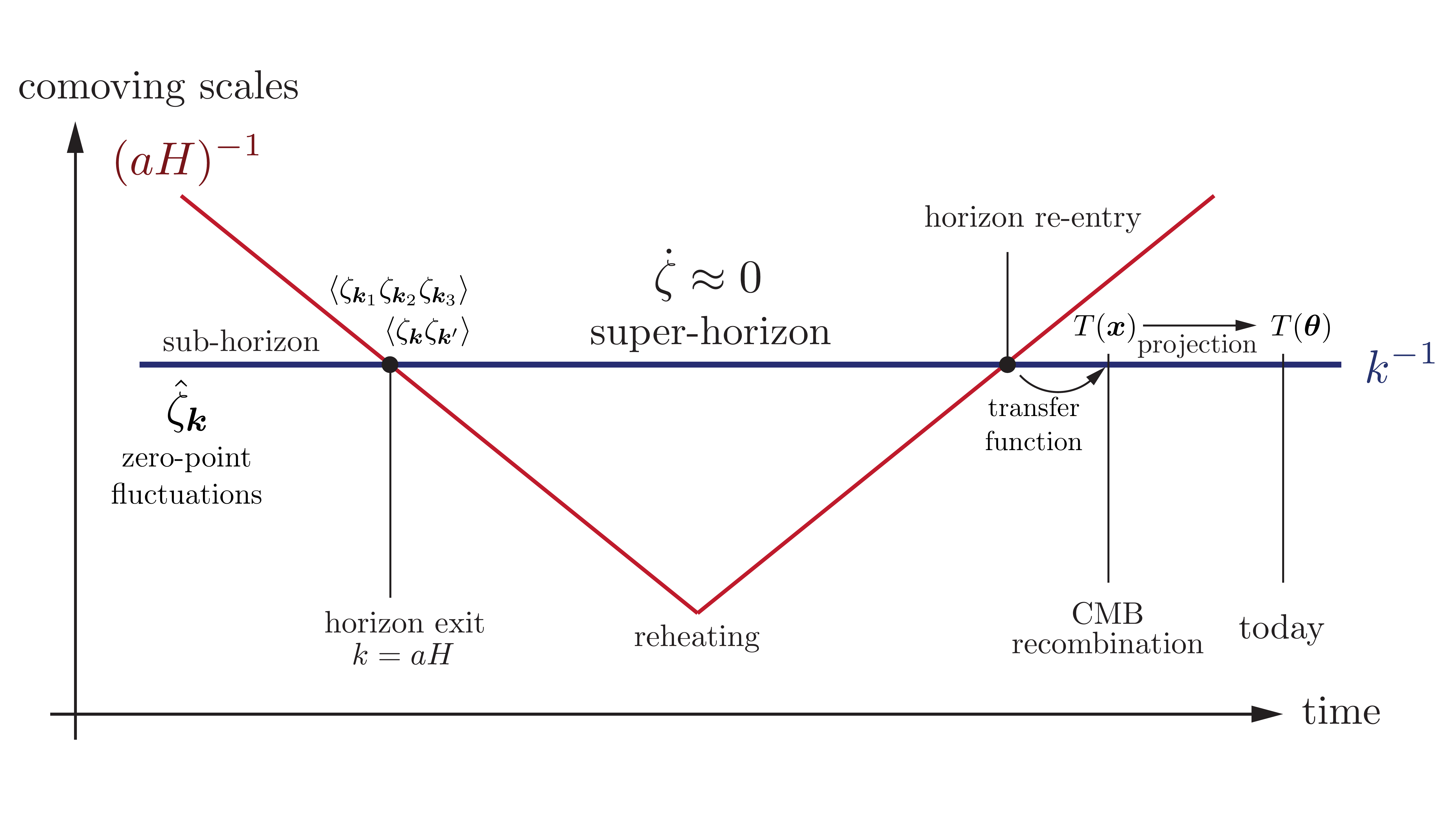}
\caption{ \footnotesize From quantum fluctuations during inflation to anisotropies in the temperature of the Cosmic
Microwave Background. 
Taken from \cite{Baumann:2009ds}.} 
\label{fig:cartoon}
\end{figure}

\noindent Let us now discuss what are the inflationary observables. 
In single-clock inflation the Fourier modes of the {comoving curvature perturbation} 
$\zeta$ and the graviton $\gamma_{ij}$, defined by 
\begin{equation}
\label{eq:efti-13}
{h}_{ij} = a^2\eu^{2\zeta} (\eu^{\gamma})_{ij} 
\end{equation} 
in the unitary gauge, are conserved as they exit the horizon, i.e.~for $k\ll aH$ \cite{Maldacena:2002vr,Weinberg:2003sw,Assassi:2012et,Senatore:2012ya}. 
They start evolving again only when they re-enter the horizon long after the end of inflation, 
during the Hot Big Bang phase (see Fig.~\ref{fig:cartoon}). Knowing $\zeta$ and $\gamma_{ij}$ 
therefore means we know the initial conditions for the growth of structure in our Universe. 
Of course, given that what we can predict are only the quantum fluctuations of $\zeta$ and $\gamma_{ij}$, 
we cannot really know the exact initial conditions. What we are interested in 
is instead the probability distribution functional of $\zeta$ and $\gamma_{ij}$. Observations suggest that these distributions are close to Gaussian. Hence, we are interested in the 
two-point correlation of $\zeta$ and $\gamma_{ij}$, or the \emph{power spectra} in Fourier space, and the deviations of their distribution 
from a Gaussian, i.e.~in \emph{primordial non-Gaussianity}.

Let us consider the correlation functions of curvature perturbation $\zeta$. Working in conformal time $\tau$ defined via
$dt = ad\tau$, which implies $\tau = -1/aH$ in exact de Sitter, we decompose $\zeta$ in Fourier modes as 
\begin{equation}
\label{eq:zeta_fourier}
\zeta(\tau,\xvec) = \int\frac{{d}^3k}{(2\pi)^3}\,\zeta_{\kvec}(\tau)\,\eu^{\iu\kvec\cdot{\xvec}}\,\,. 
\end{equation} 
The observables in the scalar sector, then, are the polyspectra 
\begin{equation}
\label{eq:polyspectra}
\begin{split}
\lim_{k\tau\to 0}\langle\zeta_{\kvec}(\tau)\zeta_{\kvec'}(\tau)\rangle' &= P_\zeta(k) \\
\lim_{\{k_i\tau\to 0\}}\langle \zeta_{\kvec_1}(\tau)\zeta_{\kvec_2}(\tau)\zeta_{\kvec_3}(\tau)\rangle' &= B_\zeta(k_1,k_2,k_3) \\ 
\lim_{\{k_i\tau\to 0\}}\langle \zeta_{\kvec_1}(\tau)\zeta_{\kvec_2}(\tau)\zeta_{\kvec_3}(\tau)\zeta_{\kvec_4}(\tau)\rangle' &= \dots\,\,,
\end{split}
\end{equation}
where $B_\zeta(k_1,k_2,k_3)$ is the \emph{bispectrum} (which depends only on the magnitude of the momenta due to 
rotational invariance), and the prime denotes that we have stripped 
a Dirac delta of momentum conservation. 

The unitary-gauge action of \eq{efti-3} is everything we need to compute not only these observables, but 
also the mixed correlation functions involving the graviton, and graviton non-Gaussianities themselves.\footnote{For 
a comprehensive study of graviton bispectra in the EFTI, see Ref.~\cite{Bordin:2020eui}.} 
However, it is when we focus on scalar correlators that the true usefulness of the EFTI becomes manifest, as we will now illustrate.

\subsection{Stueckelberg trick and decoupling limit in the EFTI} 
\label{subsec:stueckelberg_trick}

The breaking of time diffeomorphisms in the EFTI is no different in spirit from what happens in massive Yang-Mills theory, 
in which a gauge group $G$ is explicitly broken by a mass term. Now the longitudinal modes $\pi^a$ of the vector fields $A^a_\mu$ are dynamical degrees of freedom, and one can make them explicit via the so-called ``Stueckelberg trick.'' The advantage is that at high energies the $\pi^a$ are decoupled from the transverse modes of $A^a_\mu$. 
This high-energy limit is called the decoupling limit. Since in this limit the action for the $\pi^a$ is the same as what we get from a broken global symmetry group $G$,
they are often denoted as ``Goldstone bosons.'' We will use the same terminology.

To perform the Stueckelberg trick in the EFTI we need to do a broken time diffeomorphism $t = \tilde{t} + \tilde{\pi}(\tilde{x})$. 
The detailed derivation is contained in Section~3 of Ref.~\cite{Cheung:2007st}. After removing the tilde to simply the notation, the Stueckelberg trick boils down to replacing $t\to t+\pi$ 
in \eq{efti-3}. For example, we have 
\begin{equation}
\label{eq:g00_transformation}
g^{00}\to g^{00} + 2g^{0\mu}\partial_\mu\pi + g^{\mu\nu}\partial_\mu\pi\partial_\nu\pi\,\,. 
\end{equation}
Similar relations are easily derived from the standard tensor transformation rules under the broken time diffeomorphism.

Diffeomorphism invariance is restored because $\pi$ transforms nonlinearly under 
$x = \tilde{x} + \tilde{\xi}^\mu(\tilde{x})$, namely 
\begin{equation}
\label{eq:active_vs_passive_nightmare}
\tilde{\pi}(\tilde{x}) = \pi(x(\tilde{x})) + \tilde{\xi}^0(\tilde{x})\,\,. 
\end{equation}
These transformation rules can be used to find the relation between $\pi$ and $\zeta$. The precise derivation is 
contained in Appendix~A of \cite{Maldacena:2002vr} and Appendix~B of \cite{Cheung:2007sv}: one finds that 
\begin{equation}
\label{eq:zeta_pi_relation}
\zeta = {-H}\pi \big( 1 + \mathcal{O}( \varepsilon , \eta ) + \mathcal{O}( k^2 \tau^2 ) \big)  \,\, , 
\end{equation} 
where $\mathcal{O}( \varepsilon , \eta )$ represents slow-roll-suppressed terms, and $\mathcal{O}( k^2 \tau^2 )$ represents terms vanishing on superhorizon scales.
It is this relation that makes the Stueckelberg trick useful, as we will discuss next.


Now that we have reintroduced $\pi$, we can discuss what is the decoupling limit in the EFTI. Let us first write 
the metric in a way suited to the $3+1$ decomposition, i.e.~using the ADM formalism \cite{Arnowitt:1959ah,Gourgoulhon:2007ue}: 
\begin{equation}
\label{eq:efti-8}
\dif s^2 = -N^2\dif t^2 + h_{ij}(\dif x^i + N^i\dif t)(\dif x^j + N^j\dif t)\,\,. 
\end{equation} 
After reintroducing the Goldstone boson, the zero-helicity mode $\zeta$ is absent from the metric. That is, 
the spatial metric $h_{ij}$ contains only the transverse and traceless graviton. 
The quadratic mixing between $\pi$ and the metric is then a mixing between $\pi$ and the non-dynamical variables $N$ and $N^i$. 

The energy scale at which we can neglect this mixing depends on which operators are present in \eq{efti-3}, 
and which operators dominate the quadratic action. 
For example, for slow-roll inflation the mixing is $\sim\mpl^2\dot{H}\dot{\pi}\delta g^{00}$. 
After canonical normalization ($\pi_c\sim\mpl\dot{H}^{1/2}\,\dot{\pi},\delta g^{00}_c\sim\mpl\delta g^{00}$) 
we see that $E_{\rm mix}\sim\varepsilon^{1/2}H$. 
Another interesting case is when the operator $M^4_2$ gets large. The mixing is now of the form 
$\sim M^4_2\dot{\pi}\delta g^{00}$, while the canonical normalization of $\pi$ is $\pi_c\sim M^2_2\pi$, 
so that $E_{\rm mix}\sim M^2_2/\mpl$. 

Whatever $E_{\rm mix}$ is, once we are above such energy scale 
we can neglect metric fluctuations and replace \eq{efti-8} with \eq{efti-1}. The action for the Goldstone boson then simplifies to
\begin{equation}
\label{eq:efti-9}
\begin{split}
S_\pi &= \int\dif^4x\,a^3\Bigg[{-\mpl^2}\dot{H}\bigg(\dot{\pi}^2 - \frac{(\partial_i \pi)^2}{a^2}\bigg) + 2M^4_2\bigg(\dot{\pi}^2 + \dot{\pi}^3 - \frac{\dot{\pi}(\partial_i \pi)^2}{a^2}\bigg) - \frac{4}{3}M^4_3\dot{\pi}^3 + \cdots\bigg] \\
&\hp{= \int\dif^4x\,a^3\bigg[}\,\, - \frac{\bar{M}^2_2}{2}\bigg\{\frac{(1 - 2\dot{\pi})(\partial^2\pi)^2}{a^4} 
- \frac{\partial^2\pi}{a^2}\bigg(\frac{H( \partial_i \pi)^2}{a^2} 
+ \frac{4\partial_i \pi  \partial_i \dot{\pi}}{a^2}\bigg) + \cdots\bigg\}\Bigg]\,\,. 
\end{split}
\end{equation} 
The relation (\ref{eq:zeta_pi_relation}) between $\pi$ and $\zeta$ makes this action useful because of the conservation of $\zeta$ after horizon crossing. Even though the concept of the energy of a $k$-mode is not even approximately defined after horizon crossing, as long as $E_{\rm mix}\ll H$, we can use the decoupling limit action to compute $\pi$ correlators shortly after horizon crossing. These can be used to determine $\zeta$ correlators, up to $\mathcal{O}(\varepsilon,\eta)$ and $\mathcal{O}(k^2\tau^2)$ corrections.

From this discussion we see that for slow-roll inflation we always have $E_{\rm mix}/H\ll 1$, since $\varepsilon^{1/2}\ll 1$. We also see that there are no interactions in \eq{efti-9} if all EFT operators are switched off. Hence, in this case primordial non-Gaussianities come from the mixing with gravity  \cite{Maldacena:2002vr}.

Nevertheless, the decoupling-limit action is still enough to predict 
the normalization of the dimensionless power spectrum $k^3P_\zeta(k)$, 
which is proportional to $H^2/(M^2_{\rm P}\varepsilon)$, and its deviation from exact scale-invariance. Combined with limits on (or a future detection of) primordial tensor modes, whose dimensionless power spectrum is instead controlled by $H^2/M^2_{\rm P}$, we get a handle on $H$ and its time derivatives. With this we can constrain the inflaton potential $V(\phi)$. 

As another example, consider the operator $M^4_2$, which gives a quadratic action 
\begin{equation}
\label{eq:efti-10}
\int\dif^4x\,a^3\bigg[{-\frac{\mpl^2\dot{H}}{c^2_{s}}}\bigg(\dot{\pi}^2 - c^2_{s}\frac{(\partial_i \pi)^2}{a^2}\bigg)\bigg]\,\,, 
\end{equation} 
with speed of sound given by 
\begin{equation} 
\label{eq:efti-11}
\frac{1}{c^2_s} = 1-\frac{2M^4_2}{\dot{H}\mpl^2}\,\,. 
\end{equation} 
Notice that $\dot H>0$, which implies a violation of the Null Energy Condition (NEC), is no longer associated to a ghost-like instability if $M_2$ is sufficiently large. Here, the general connection between the violation of the NEC and instabilities  \cite{Hsu:2004vr,Dubovsky:2005xd} persists because there is a gradient instability in the model. However, from a bottom-up point of view, it is possible to construct a stable EFT (called Ghost Condensate) that allows $\dot H>0$ \cite{Creminelli:2006xe}. 

It is not clear if NEC violating models can be UV completed, and indeed there are results that suggest otherwise \cite{Hartman:2016lgu}. Below, we will focus on the $\dot H <0$ case, where one finds 
\begin{equation}
\label{eq:no_cs_determination}
k^3P_\zeta(k)\propto\frac{H^2}{M^2_{\rm P}\varepsilon c_{s}}\,\,. 
\end{equation}
Even if we detect primordial tensor modes, we still cannot disentangle between $\varepsilon$ and the speed of sound. To get a handle on $c_{s}$ in this case, we need to look at interactions. Indeed, 
turning on the operators $M^4_2, M^4_3$ results in 
\begin{equation}
\label{eq:efti-12}
S^{(3)}_\pi = \int\dif^4x\,a^3\bigg[
\frac{M_{\rm P}^2\dot{H}}{c_{s}^2}(1-c_{s}^2)
\left(\frac{\dot{\pi}(\partial_i \pi)^2}{a^2}-
\left(1+\frac{2}{3}\frac{\tilde{c}_3}{c_{s}^2}\right)\dot{\pi}^3
\right)
\bigg]\,\,, 
\end{equation} 
where 
\begin{equation}
\label{eq:ctilde3}
\frac{4}{3}\frac{M^4_3}{{\dot{H}\mpl^2}} = \frac{2}{3} \frac{1-c^2_s}{c^2_s} \frac{\tilde{c}_3}{c^2_s}\,\,. 
\end{equation} 
A speed of sound different from $1$ means that there is a specific interaction $\dot{\pi}(\partial_i \pi)^2$ 
uniquely determined by $c^2_s$. Hence, by constraining the non-Gaussianities generated by $\dot{\pi}(\partial_i \pi)^2$, we can constrain the speed of propagation of the scalar mode in single-clock inflation. 

This relation between the quadratic action and interactions is forced by the nonlinear realization of time diffeomorphisms. In the decoupling limit, this symmetry is reduced to invariance under de Sitter dilations and boosts, which are generate by 
\begin{subequations}
\label{eq:boosts-3}
\begin{align}
{\xi}^\mu_0 &= {-H^{-1}\delta^\mu_0} + x^i\delta^\mu_i\,\,, \label{eq:boosts-3-1} \\
{\xi}^\mu_i &= \delta^\mu_i x^2 - 2 x^i x^j\delta^\mu_j + 2H^{-1}x^i\delta^\mu_0\,\,, \label{eq:boosts-3-2}
\end{align}
\end{subequations} 
where $x^2 = -\eta^2 + \xvec^2$. Under these, $\pi$ transforms nonlinearly \cite{Creminelli:2012ed}: for infinitesimal transformation parameter $\lambda^a$, with $a\in \{0,1,2,3\}$, we have 
\begin{equation}
\label{eq:boosts-2}
\delta\pi = \lambda^a\xi^\mu_a\partial_\mu\pi + \lambda^a{\xi}^0_a\,\,,
\end{equation}
where we used \eq{active_vs_passive_nightmare}.
The action is invariant under \eq{boosts-3-2} only if the coefficient of the $\dot{\pi}(\partial_i \pi)^2$ operator in \eq{efti-12} has 
that specific dependence on the speed of sound.

Finally, let us discuss in more detail the decoupling limit for the case where the operators $M^4_n$ are turned on. 
In the case of $c^2_s\ll 1$, we have $E_{\rm mix}/H\sim (\varepsilon/c^2_s)^{1/2}$. So we see that 
at a fixed $\varepsilon$ the speed of sound cannot be too small if we want to use the decoupling-limit action of \eq{efti-9}.\footnote{The current bound on the speed of sound (which comes from constraints on primordial non-Gaussianity, 
as we will discuss in a moment) and on $\varepsilon$ (from the absence of detection of primordial $B$-mode polarization of the CMB) are $c^2_s\gtrsim 4\times10^{-4}$ \cite{Akrami:2019izv} and $\varepsilon\lesssim 4\times10^{-3}$ at $95\%\,{\rm CL}$ \cite{Planck:2018jri}. The region of parameter space for which $E_{\rm mix}/H\ll 1$ is still allowed: however, in case of a detection of these two parameters it could prove necessary to go beyond the decoupling limit to compute accurately the correlators of $\zeta$.} 
Given that the mixing scale depends on which operators one is considering, a similar analysis must be carried out 
depending on which kind of interactions of $\pi$ one wants to constrain.

\subsection{EFT cutoff} 
\label{subsec:cutoff_scales}

\noindent One might wonder why we delayed the discussion of the cutoff scale, a crucial concept for any effective theory, 
up to now. The reason is that the cutoff of the EFTI depends on what operators dominate the action, so it was necessary to 
first introduce these operators. In this paper we review only what happens in $P(X,\phi)$ theories for $c^2_s\ll 1$, 
referring to \cite{Cheung:2007st} for more details on other cases.

We focus on the operator $\dot{\pi}(\partial_i \pi)^2$ 
in \eq{efti-12} whose coefficient is fixed by $c_s$. We can estimate the UV cutoff of the theory by working in the subhorizon limit and finding the 
maximum energy at which the tree-level scattering of Goldstones is perturbative. The calculation is straightforward, the only complication coming from the non-relativistic dispersion relation $\omega = c_{s} k$. The cutoff (or ``strong-coupling'') scale $\Lambda_\star$ turns out to be 
\begin{equation}
\label{eq:cutoffs-1} 
\Lambda^4_\star \simeq \frac{f^4_\pi c^7_s}{1-c^2_s}\,\,, 
\end{equation} 
where $f^4_\pi$ is defined based on the kinetic term \eq{efti-10}
\begin{equation}
\label{eq:cutoffs-2}
f^4_\pi = \frac{2\mpl^2\abs{\dot{H}}}{c^2_s}\,\,. 
\end{equation} 
The scale $\Lambda_\star$ indicates the energy at which infinitely many EFT operators become important. So the effective description breaks down and new physics must come into the game \cite{Cheung:2007st,Baumann:2011su}. 

In this example, our construction of the EFT for the fluctuations was motivated by the $P(X,\phi)$ model. However, only when $\Lambda_\star\gg f_\pi$ can we think of $P(X,\phi)$ as a UV completion of the EFTI: a weakly coupled ``effective field theory for inflation'' that interpolates between the trivial background $X = 0$ and the rolling background $X\neq 0$ \cite{Creminelli:2003iq,Weinberg:2008hq}. When $c_s\ll 1$, or more generally $\Lambda_\star\ll f_\pi$, one needs an alternative since $P(X,\phi)$ is strongly coupled around $X=0$.
In the next section, we will discuss how this observation provides a theoretically-motivated target for primordial non-Gaussianity.

\subsection{Amplitude and shape of primordial non-Gaussianity}
\label{subsec:PNG_shapes}

\noindent The chief observable that describes deviations from Gaussianity is the bispectrum, 
that we introduced in Section~\ref{subsec:observables}. The conservation of $\zeta$ at super-horizon scales implies that at leading order in slow-roll approximation $B_\zeta(k_1,k_2,k_3)$ is scale-invariant. 
By momentum conservation the three momenta in the bispectrum form a triangle. It is useful to factor out an amplitude and introduce a function that describes the dependence on the shape of triangle:
\begin{equation}
\label{eq:efti-14}
B_\zeta(k_1,k_2,k_3) = \frac{18}{5}f_{\rm NL}\Delta^4_\zeta\frac{S(k_1,k_2,k_3)}{k_1^2k_2^2k_3^2}\,\,,
\end{equation}
where $\Delta^2_\zeta = k^3P_\zeta(k)$, the factor of $18/5$ is a historical convention, 
and the dimensionless shape function $S(k_1,k_2,k_3)$ is normalized to $1$ in the \emph{equilateral} 
configuration, $S(k,k,k) = 1$.

\begin{figure}[t]
\centering
\hspace{-.5in}\includegraphics[width=0.7\textwidth]{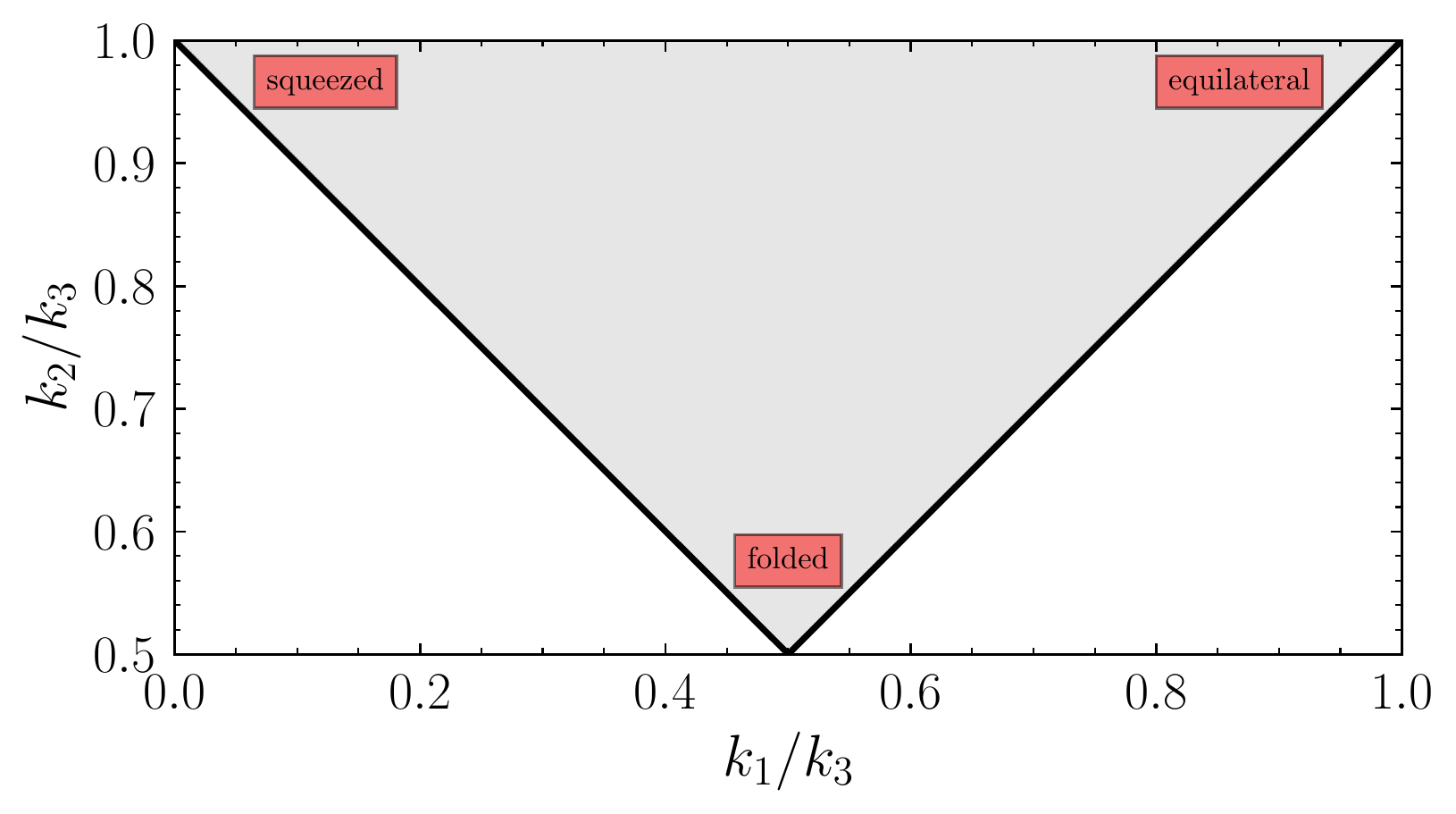}
\caption{ \footnotesize Range of momenta $\cal V$ for plotting a shape function $S(x_1,x_2,1) = {\cal S}(x_1,x_2)$, 
and for computing the cosine between two shapes. The \emph{equilateral} configuration is 
$x_1\to1,x_2\to 1$, while the \emph{squeezed} configuration is $x_1\to 0,x_2\to 1$, i.e.~it is 
the limit in which one of the modes ($k_1$) becomes much longer than the other two. 
The configuration $x_1\to1/2,x_2\to1/2$ is called \emph{folded}, and it corresponds to very squashed isosceles triangle.} 
\label{fig:shapes_plot-1} 
\end{figure}

\begin{figure}[t]
\centering \hspace{-0.25in}
\begin{tabular}{cc}
\includegraphics[width=0.49\columnwidth]{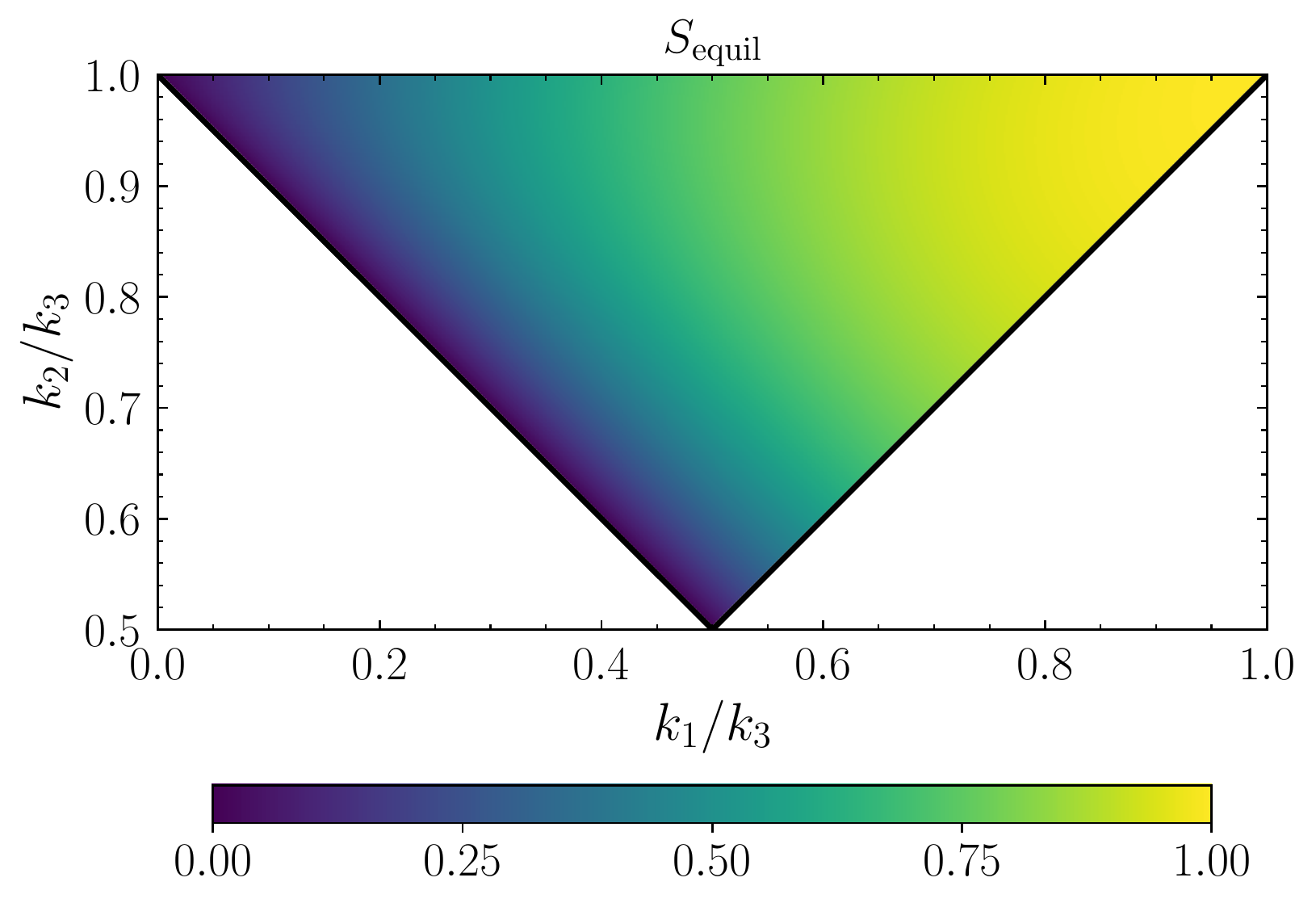} 
\includegraphics[width=0.49\columnwidth]{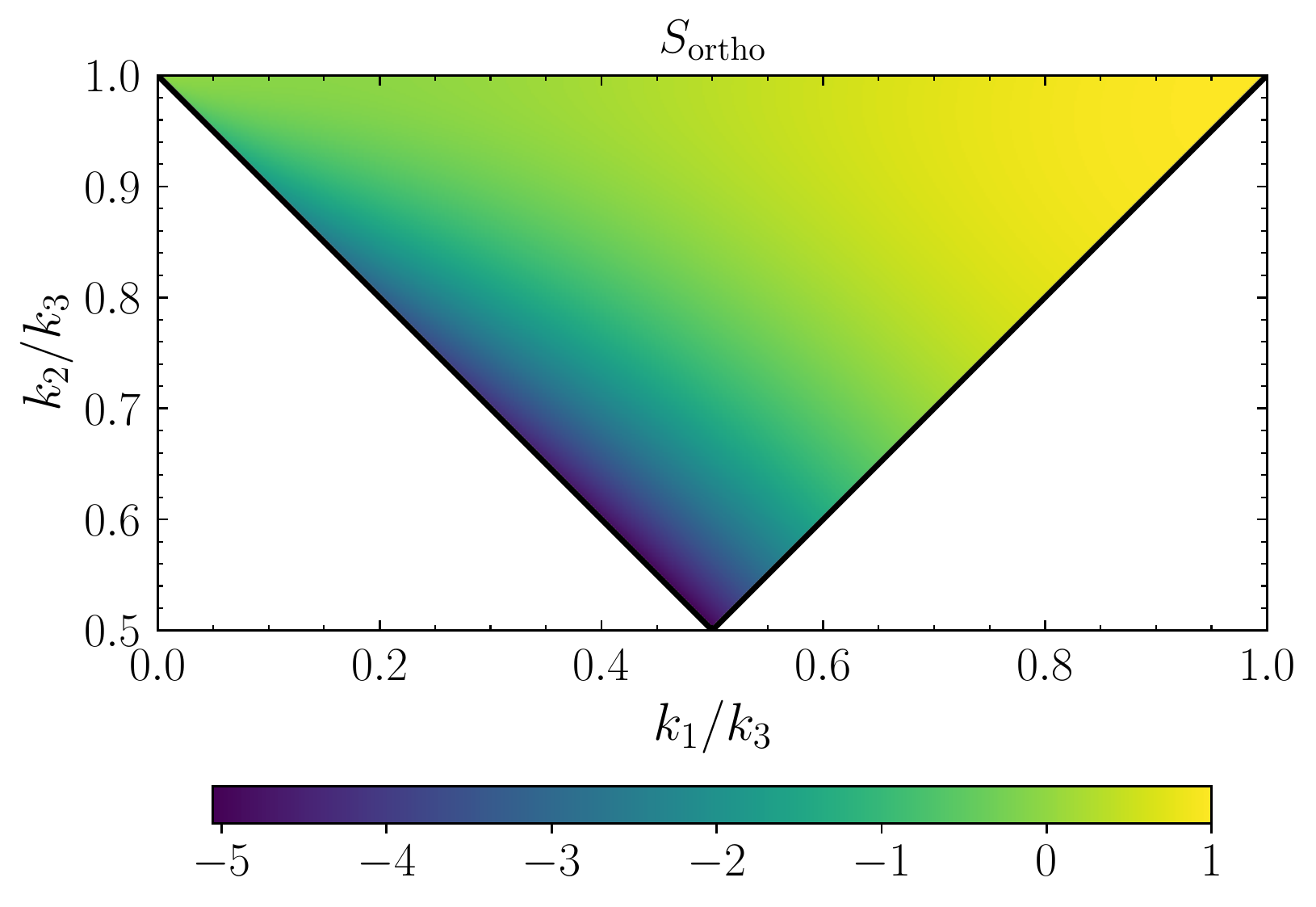}
\end{tabular}
\caption{\footnotesize {Left panel} -- Equilateral template of \eq{efti-21}, which peaks in the equilateral configuration. 
{Right panel} -- Orthogonal template. We see that it peaks in the 
folded configuration.}
\label{fig:shapes_plot-2}
\end{figure}

The amplitude $\Delta^2_\zeta$ of the primordial power spectrum is very well measured by the \emph{Planck} satellite 
($\smash{\Delta^2_\zeta \approx 4.1\times10^{-8}}$). Hence the overall level of primordial non-Gaussianity is controlled by the 
parameter $f_{\rm NL}$. One can estimate $f_{\rm NL}$ by comparing the quadratic and cubic Lagrangians of the Goldstone boson as \cite{Cheung:2007st} 
\begin{equation}
\label{eq:efti-15}
\frac{{\cal L}^{(3)}_\pi}{{\cal L}^{(2)}_\pi}\sim f_{\rm NL}\zeta\,\,, 
\end{equation} 
where derivatives are estimated by evaluating them at horizon crossing. 
Let us look, for example, at \eq{efti-10} and the interactions of \eq{efti-12}. At crossing, we have $\partial_0\sim H$, but $\partial_i/a\sim H/c_{s}$. 
Therefore the operator with more spatial derivatives in \eq{efti-12} is enhanced when $c^2_s\ll 1$. We have 
\begin{equation}
\label{eq:efti-16}
\frac{{\cal L}^{(3)}_\pi}{{\cal L}^{(2)}_\pi}\sim\frac{H\pi \big(\frac{H}{c^2_s}\pi\big)^2}{(H\pi)^2} 
\sim\frac{H\pi}{c^2_s}\sim\frac{\zeta}{c^2_s}\,\,. 
\end{equation} 
Hence $f_{\rm NL}\sim 1/c^2_s$. An exact computation gives \cite{Gruzinov:2004jx,Chen:2006nt,Senatore:2009gt}
\begin{equation}
\label{eq:fNLdotpinablapisquared}
f_{\rm NL}^{\dot{\pi}(\partial_i \pi)^2} = \frac{85}{324}\bigg(1-\frac{1}{c^2_s}\bigg)\,\,.
\end{equation}
This non-Gaussianity is peaked near the equilateral configuration. This is because derivatives of $\pi$ decay fast outside the horizon, and little contribution comes from the period when the modes or deep inside the horizon due to their fast oscillations. So the interaction is maximal when all three modes cross the horizon around the same time. From the point of view of data analysis and the actual detection of primordial non-Gaussianity it is useful to find a template that 
is easy to manipulate while still being a good representation of the bispectrum shape of the $\dot{\pi}(\partial_i\pi)^2$ operator. 
For example, it is very useful to have a template that is separable in $k_1,k_2,k_3$. Finding such templates 
can be achieved by the introduction of the \emph{cosine} between two shapes. First, because of scale invariance we can 
always rewrite the shape function as 
\begin{equation}
\label{eq:efti-18}
S\bigg(\frac{k_1}{k_3},\frac{k_2}{k_3},1\bigg) \equiv S(x_1,x_2,1) \equiv {\cal S}(x_1,x_2)\,\,. 
\end{equation}
Organizing the momenta as $k_1\leq k_2\leq k_3$, the conservation of momentum amounts to requiring 
$0\leq x_1\leq 1$ and $1-x_1\leq x_2\leq 1$: this region $\cal V$ is shown in Fig.~\ref{fig:shapes_plot-1}. 
Given two shapes ${\cal S}_1,{\cal S}_2$, one can check that the integral 
\begin{equation}
\label{eq:efti-19}
{\cal S}_1\cdot{\cal S}_2 = \int_{\cal V}\dif x_1\dif x_2\,{\cal S}_1(x_1,x_2)\,{\cal S}_2(x_1,x_2) 
\end{equation} 
defines a scalar product. Then, the cosine 
\begin{equation}
\label{eq:efti-20}
\frac{{\cal S}_1\cdot{\cal S}_2}{\sqrt{({\cal S}_1\cdot{\cal S}_1)({\cal S}_2\cdot{\cal S}_2)}} 
\end{equation} 
quantifies how much two shapes are similar \cite{Babich:2004gb,Senatore:2009gt}. It is possible to check that the shape 
\begin{equation}
\label{eq:efti-21} 
S_{\rm equil}(k_1,k_2,k_3) \propto \bigg(\frac{k_1}{k_2} + \text{$5$ perms.}\bigg) - \bigg(\frac{k^2_1}{k_2k_3} + \text{$2$ perms.}\bigg) - 2 
\end{equation}
has strong overlap with the shape of $\dot{\pi}(\partial_i\pi)^2$. This is the \emph{equilateral template}, which we plot in Fig.~\ref{fig:shapes_plot-2}.

The cosine was originally introduced to study how much two different templates 
could be distinguished in CMB or large-scale structure data. Given a bispectrum shape, 
one can build an optimal estimator for its $f_{\rm NL}$. Then, if two shapes have a 
small scalar product, the optimal estimator for one shape will be vary bad in 
detecting non-Gaussianities coming from the other, and vice versa (see Ref.~\cite{Babich:2004gb} for more details). 
As such, one could modify the definition of cosine to account for the noise and window function of a given 
CMB or large-scale structure experiment. 

For a comprehensive search, we need a basis of shapes onto which one can project the bispectra of the EFTI \cite{Senatore:2009gt}. 
For $P(X,\phi)$ theories these are the bispectra from $\dot{\pi}(\partial_i \pi)^2$ 
and from $\dot{\pi}^3$, whose $f_{\rm NL}$ is 
\begin{equation}
\label{eq:fNLdotpicubed} 
f_{\rm NL}^{\dot{\pi}^3} = 
{\frac{10}{243}}\bigg(1-\frac{1}{c^2_s}\bigg)\bigg(\tilde{c}_3+\frac{3}{2}c^2_s\bigg)\,\,. 
\end{equation} 
The \emph{orthogonal} shape was introduced for the purpose of obtaining this projection. We plot it in the right panel of Fig.~\ref{fig:shapes_plot-2}.

The orthogonal template takes its name from the fact that it has zero overlap with the equilateral one, 
${\cal S}_{\rm equil}\cdot{\cal S}_{\rm ortho} = 0$. So we can think of it as a second basis 
vector in the infinite-dimensional space of shapes. Via the cosine we can then obtain 
$\smash{f^{\dot{\pi}(\partial_i \pi)^2}_{\rm NL}}$ and $f^{\dot{\pi}^3}_{\rm NL}$ in terms of $f_{\rm NL}^{\rm equil}$ and $f_{\rm NL}^{\rm ortho}$.

Constraints on $f_{\rm NL}^{\rm equil}$ and $f_{\rm NL}^{\rm ortho}$ can then be translated into constraints on the speed of sound and the $\tilde{c}_{3}$ parameter using \eqsII{fNLdotpinablapisquared}{fNLdotpicubed}. This is what has been done with CMB data from the 
WMAP and \emph{Planck} satellites \cite{Akrami:2019izv}, and recently from large-scale structure data 
from the BOSS galaxy survey (see discussion and references in Section~\ref{datasec}). 

Before concluding this section, let us point out a theoretically motivated target for 
$\smash{f_{\rm NL}^{\rm equil}}$ and $\smash{f_{\rm NL}^{\rm ortho}}$, following Refs.~\cite{Creminelli:2003iq,Baumann:2011su,Alvarez:2014vva}. These observables are directly related to the cutoff of the EFT: larger $f_{\rm NL}$ means lower strong coupling scale. As discussed in section \ref{subsec:cutoff_scales}, the UV completion of the EFTI has a qualitatively different flavor when $\Lambda_\star\ll f_\pi$, and in particular when $c_s\ll 1$. Given the predictions \eq{fNLdotpinablapisquared} and \eq{fNLdotpicubed} we see that a natural target is $\smash{f_{\rm NL}^{\rm equil},f_{\rm NL}^{\rm ortho}\sim 1}$.
Much effort has been, and continues to be, devoted to reaching this target.

\subsection{Local non-Gaussianity and consistency relations}
\label{subsec:CRs}

\begin{figure}[t]
\centering
\includegraphics[width=0.7\textwidth]{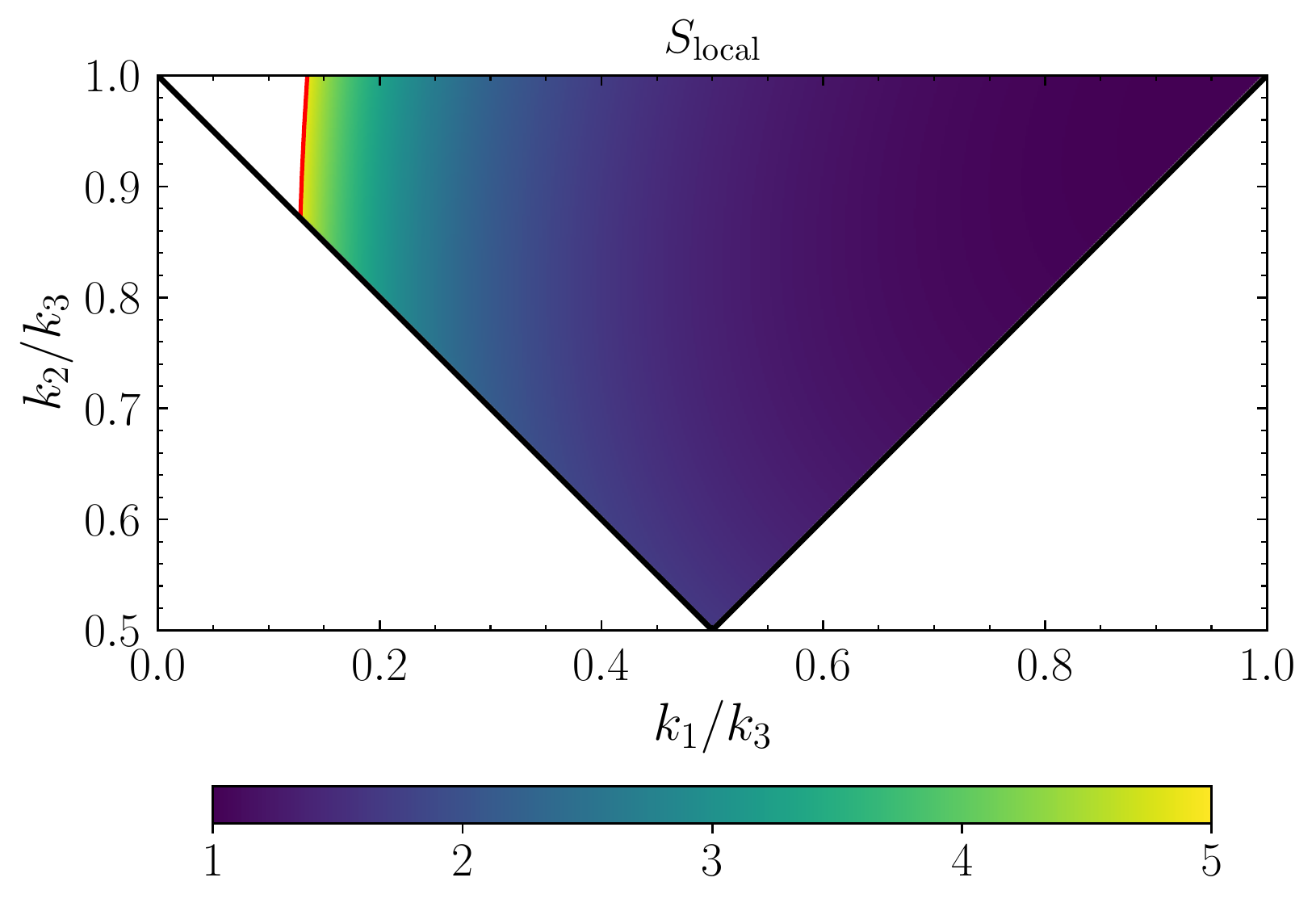}
\caption{ \footnotesize Local template. It diverges as $1/x_1$ in the squeezed limit.} 
\label{fig:shapes_plot-5}
\end{figure}

\noindent There is another important type of non-Gaussianity, the so-called 
non-Gaussianity of the \emph{local} type. Its template is 
\begin{equation}
\label{eq:efti-24}
S_{\rm local}(k_1,k_2,k_3) = \frac{1}{3}\frac{k^2_1}{k_2k_3} + \text{$2$ perms.} 
\end{equation} 
It derives its name from the fact that it comes from a nonlinear local correction to 
a Gaussian variable: $\zeta(\xvec)\to\zeta(\xvec) + \zeta^2(\xvec)$. 
Importantly, in Fig.~\ref{fig:shapes_plot-5} we see that it peaks in the squeezed configuration, 
so it is very distinguishable from equilateral and orthogonal non-Gaussianity.

Local non-Gaussianity vanishes in single-clock inflation. 
This is a consequence of the fact that in single-clock inflation the squeezed limit of the bispectrum is uniquely
fixed in terms of the power spectrum by the \emph{consistency relation} \cite{Maldacena:2002vr,Creminelli:2004yq,Cheung:2007sv}: 
\begin{equation}
\label{eq:efti-25}
\lim_{k_1\to 0}B_\zeta(k_1,k_2,k_3) = \bigg[{-\frac{\dif\ln k^3_SP_\zeta(k_S)}{\dif\ln k_S}} 
+ {\cal O}\bigg(\frac{k^2_L}{k^2_S}\bigg)\bigg]P_\zeta(k_L)P_\zeta(k_S) 
\,\,, 
\end{equation} 
where we have defined 
\begin{equation}
\label{eq:efti-26}
\kvec_L = \kvec_2+\kvec_3\,\,,\quad\kvec_S = \frac{\kvec_2-\kvec_3}{2}\,\,. 
\end{equation}
This result can be derived in the following way. When the long mode $k_1$ goes outside the horizon, 
the associated perturbations in ADM variables $N$ and $N^i$ defined in $\zeta$ gauge go to zero. 
In this limit the metric becomes 
\begin{equation}
\label{eq:efti-27}
\dif s^2 \approx {-\dif t}^2 + a^2(t)\eu^{2\zeta(\xvec)}\delta_{ij}\dif x^i\dif x^j\,\,, 
\end{equation} 
where we used that $\zeta$ becomes a constant. 
The evolution of short-wavelength modes is then the same as in an unperturbed FLRW Universe, but with a \emph{local} 
scale factor $a^2(t)\eu^{2\zeta(\xvec)}$. Hence the correlation between the long mode and two short modes is given by a scale tranformation as in \eq{efti-25}. Then, we can say that single-clock inflation does \emph{not} produce local non-Gaussianity: if we measure 
the squeezed limit of a three-point function, we do not learn anything more than what we would learn from 
measuring the two-point function. In other words, the long-wavelength field~$\zeta(\xvec)$ can be locally removed from the metric by a large gauge transformation given by \eq{efti-25}, and therefore it is locally unobservable.

Consistency relations for single-field inflation have been an active area of study in the past decade. The main result given in \eq{efti-25} has been generalized for higher-order correlation functions and including soft tensor modes~\cite{Creminelli:2012ed,Hinterbichler:2012nm,Hinterbichler:2013dpa,Senatore:2012wy,Assassi:2012zq}, as well as the case of multiple soft limits~\cite{Mirbabayi:2014zpa,Joyce:2014aqa}.

The same line of reasoning used to derive inflationary consistency relations can be extended from the horizon exit 
to re-entry of the long modes. This results in consistency relations for \emph{cosmological observables}, that 
all take the form of \eq{efti-25}. The more famous are the so-called consistency relations for large-scale structure 
\cite{Peloso:2013zw, Kehagias:2013yd, Creminelli:2013mca,Valageas:2013cma,Creminelli:2013poa,Creminelli:2013nua,Kehagias:2013rpa}, but similar results have been obtained for CMB anisotropies and CMB spectral distortions \cite{Creminelli:2011sq,Mirbabayi:2014hda,Cabass:2018jgj}. 

At phenomenological level, consistency relations play a very important role.
They imply that any detection of local non-Gaussianity, i.e.~a violation of the consistency relations, 
would rule out all models described by the single-field EFTI. This is why most of the current experimental effort, as far as the physics of the primordial Universe is concerned, is focused on local non-Gaussianity.

\subsection{Beyond single-clock inflation}
\label{subsec:beyond_SF}
\noindent Presence of additional degrees of freedom can significantly modify the predictions of inflation. For instance, inflationary models with extra massless fields (often called ``multifield models'') can generate local non-Gaussianity \cite{Bartolo:2001cw,Bernardeau:2002jy,Lyth:2002my,Dvali:2003em,Zaldarriaga:2003my}. Therefore, they can violate the single-field consistency condition \eq{efti-25}. 

In fact, new degrees of freedom even if massive leave their imprints in the squeezed limit of non-Gaussian correlators \cite{Chen:2009zp,Chen:2010xka,Chen:2012ge,Arkani-Hamed:2015bza}. For instance, the exchange of a scalar field of mass $m$ leads to a squeezed-limit behavior proportional to 
\begin{equation}
\label{eq:efti-29}
\text{$\frac{1}{k^3_Lk^3_S}\bigg(\frac{k_L}{k_S}\bigg)^{\frac{3}{2}+\iu\mu} + \text{c.c.}$ \quad with \quad $i\mu =\sqrt{  \frac{9}{4}-\frac{m^2}{H^2}} \,\,.$} 
\end{equation} 
This becomes similar to the local non-Gaussianity as $m^2/H^2\to 0$, while there is a distinct oscillatory behavior when $m/H>3/2$. If the exchanged particle had a nonzero spin $s$, then the squeezed limit correlator would also have an angular dependence $P_s(\vers{k}_L\cdot\vers{k}_S)$. Hence by investigating the squeezed-limit of inflationary correlators, we are effectively doing spectroscopy of that era. 

Of course, these additional degrees of freedom can be included in the EFTI to reproduce the known results, but more importantly to explore the full range of possibilities that are consistent with the symmetries. Some works along these lines are \cite{Senatore:2010wk,Baumann:2011nk,Mirbabayi:2015hva,Lee:2016vti,Delacretaz:2016nhw,Bordin:2018pca}. 

To highlight an example, let us recall the Higuchi bound: massive unitary representations of de Sitter group with nonzero spin cannot be arbitrarily light: $m^2 > s(s-1)H^2$ \cite{Higuchi:1986py}. This bound suppresses the strength of the squeezed-limit signal coming from spinning degrees of freedom (see \eq{efti-29}). However, even if inflationary spacetime is very close to de Sitter, the Higuchi bound can be strongly violated because dS isometries are broken during inflation (there are preferred time slices). EFT of Inflation helps systematically study this possibility, which indeed leads to phenomenologically interesting signatures \cite{Bordin:2018pca}.

Constraints on primordial non-Gaussianity have been so far driven by CMB experiments, but 
to improve these constraints we should explore other probes. 
Chief among these probes is large-scale structure. In order to obtain robust constraints on primordial non-Gaussianity 
from large-scale structure, however, we must have an accurate theoretical description of nonlinearities 
from gravitational collapse, since these act as a ``noise'' for the extraction of the primordial signal. 
This is where the EFT of LSS, 
another success in the application of effective field theory techniques to cosmology, comes into play.

\section{Effective Field Theory of Large-Scale Structure}\label{eftlsssec}

The density fluctuations seeded during inflation can be 
observed through  
perturbations of the CMB and large-scale
structure. Large-scale structure is the distribution
of matter on cosmological scales at low redshifts. This distribution
is measured through various channels: 
weak lensing of the CMB and galaxies, 
spectroscopic galaxy surveys,  
Lyman-$\alpha$ intensity absorption patterns etc.
In order to get more information about our Universe one has 
to establish the connection between these observables 
and fundamental properties of the Universe.  
To that end, it is desirable to analyze large-scale structure data just like
the CMB, 
where one uses linear cosmological
perturbation theory to extract cosmological parameters from the observed 
spectra of 
temperature and polarization fluctuations.
However, the large-scale structure observables are somewhat different 
from the CMB ones. The low-redshift Universe is strongly affected by
gravitational instability and complex galaxy formation physics, neither of which can be adequately modeled within
linear cosmological perturbation theory. 
On the other hand, the number of modes 
available for measurements in large-scale structure
experiments 
is nominally much larger than that of the CMB 
because the matter distribution is essentially three-dimensional. 
Potentially, this may lead to very precise measurements of 
cosmological parameters provided that large-scale structure can be accurately modeled. 

The effective field theory of large scale structure \cite{Baumann:2010tm,Carrasco:2012cv} and its spin-offs \cite{Porto:2013qua,Blas:2015qsi,Vlah:2015sea} are theoretical tools for accurate analytic 
calculations of non-linear structure formation
in our Universe. The main object of this theory are small fluctuations in the number density of biased tracers, such as galaxies, expanded around homogeneous and isotropic background given by a cosmological model at hand. The cutoff of this theory is given by the scale where the gravitational collapse become very nonlinear or where the impact of astrophysical processes involving baryons is significant. Below this cutoff, the evolution and interactions of the long-wavelength density fluctuations are fixed by gravity as the only long-range force and symmetries of the system. Remarkably, this allows for the description of structure formation on large scales in terms of a weakly-coupled theory, even when the details of complicated baryonic physics governing galaxy formation are unknown. In this way the EFT of LSS provides a direct link between the (non)-Gaussian initial conditions set by inflation and the late Universe observables. In what follows we will review
the current state of this field.

\subsection{Fluid description of the large-scale structure}


In order to illustrate the main principles of the EFT of LSS, we will focus on a simple example where 
the Universe is dominated by
collisionless non-relativistic particles. Such example is already very generic. These particles can represent dark matter, small dark matter halos (as is often the case in numerical N-body simulations) or they can be any other compact objects, such as primordial black holes.
Since the gravitation collapse takes place sufficiently
inside the Hubble horizon, it essentially occurs in the Newtonian
non-relativistic regime. In this regime, the exact description of a system of $N$ identical
particles of mass $m$ which interact only gravitationally   
is given by the Vlasov equation for the total phase-space 
probability
distribution function (PDF) $f(t,\p,\x)$,
\be
\frac{\d f}{\d t} +\frac{p^i}{m a^2}\frac{\d f}{\d x^i}-m
\sum_{a,b; a\neq b}
\frac{\d \phi_a}{\d x^i}\frac{\d f_b}{\d p^i}=0\,,
\ee
where $\phi_a$ and $f_b$ are the single-particle gravitational potentials 
and phase-space densities, and $f=\sum_{a=1}^N f_a$.
This setup allows us to obtain the equation of motion for 
the long-wavelength degrees of freedom by explicitly integrating out the UV modes. This is in practice achieved by coarse-graining the Boltzmann
equation by means of a low-pass filter with some cutoff scale $\Lambda$ 
and taking the first two moments of the resulting filtered 
phase-space PDF, which yields~\cite{Baumann:2010tm,Carrasco:2012cv} 
\be
\label{eq:fluid}
\begin{split}
& \d_\tau \delta +\d_i[(1+\delta)v^i]=0\,,\\
& \d_\tau v^i + \mathcal{H} v^i+\d^i\Phi +v^j \d_j v^i=-\frac{1}{a\rho}\d_j\tau^{ij}\,,\\
& \Delta \Phi = \frac{3}{2}\mathcal{H}^2 \Omega_m \delta\,.
\end{split} 
\ee
In these equations we used conformal time $d\tau=dt/a$ , $\Omega_m$ is the time-dependent
matter density fraction which enters the Friedmann equation, $\mathcal{H}=\d_\tau a/a$ is the conformal Hubble parameter, 
$\Phi$ is the gravitational potential and $\delta\equiv \delta \rho/\bar{\rho}$ and  $v^i$ are the filtered density contrast and 
peculiar velocity fields, constructed by coarse-graining 
the density and momentum fields,
\be 
\begin{split}
& \rho \equiv \frac{m}{a^3}\int d^3p~f(\p,\x)\,,\quad \frac{v^i}{\rho}\equiv\frac{1}{a^4}\int d^3p~p^i f(\p,\x)\,.
\end{split}
\ee 
As argued in~\cite{Baumann:2010tm}, consistent truncation of the infinite hierarchy of moments of the Boltzmann equation
is possible as long as the scales of interest are larger than the 
effective mean free path of dark matter particles.
Crucially, on the right hand side of the Euler equation in~\eqref{eq:fluid} we see the appearance of
an effective stress-tensor $\tau^{ij}$, which is generated by integrating out the short-scale
fluctuations. As we will argue shortly, this effective stress-tensor can be expanded in the powers of spacial derivatives and long-wavelength density fields on large scales. Therefore, the fluid description of our Universe is possible as long as the following
condition is satisfied
\be
\frac{k}{k_{\rm NL}} \ll 1\,,
\ee
where $k$ is a wavenumber of density perturbations and
$k^{-1}_{\rm NL}\sim 5~\text{Mpc}$ (at redshift zero) is the so-called nonlinear scale for which the variance of the density field becomes of order unity:
$(2\pi^2)^{-1}P_{\rm lin}(k_{\rm NL})k^3_{\rm NL}\approx 1$. 

Eq.~\eqref{eq:fluid} 
is the equation of motion for the 
long-wavelength degrees of freedom. We have obtained it starting from a simple exact description of a self-gravitating system and explicitly integrating out the UV modes. However, as in any other EFT, the same equations of motion can be derived identifying the relevant long-wavelength degrees of freedom and imposing all symmetries of the system~\cite{Mercolli:2013bsa}, even when the UV model is unknown. 
Therefore, the long-wavelength description given by Eq.~\eqref{eq:fluid} is 
universal, i.e.~by construction it covers all 
possible microscopic scenarios 
of structure formation.
This description allows one to capture 
effects of unspecified UV physics in a systematic 
and robust fashion. 
This is not surprising, given that the EFT decoupling 
principle guarantees that the impact of any UV physics can be captured
by effective operators constructed from the long-wavelength 
degrees of freedom only.

\subsection{Stress tensor and (non--)locality in time} 

Filtering short-scale modes produces an effective stress-energy tensor 
in the Euler equation \eqn{eq:fluid}. This tensor depends only on the long-wavelength
degrees of freedom, i.e.~the smoothed density contrast and peculiar
velocity. These two fields contain deterministic and stochastic 
components. The deterministic component is correlated with the 
long-wavelength fields, while the stochastic is not. However, its statistical 
properties are strongly constrained by symmetries, i.e.~the presence of the stochastic component still allows the 
theory to be predictive.

On sufficiently large scales the non-linear evolution is negligible,
and hence these quantities are small. Thus, the deterministic part of the effective stress-tensor
can be Taylor-expanded in powers of the wavenumbers and the large scale 
fields and spatial derivatives. 
The most general expression consistent with the rotation invariance
and the equivalence principle is given by~\cite{Carrasco:2013mua}
\be 
\label{eq:taunonloc}
\frac{1}{a\rho}\d_j\tau^{ij}=\int d\tau'~K(\tau,\tau')
~\d^i\delta(\x_{\rm fl}[\x,\tau;\tau'],\tau')+...\,,
\ee
where $K(\tau,\tau')$ is a time propagator, $\x_{\rm fl}[\x,\tau;\tau']$
is the position of the fluid element $(\x,\tau)$ at time $\tau'$.
We emphasize that the effective stress tensor depends on fields evaluated
on the past light-cone, i.e.~the EFT of LSS is in general nonlocal in time \cite{Carroll:2013oxa}. 
In conventional effective field theories the time scale of short
modes is faster than the time scale of long-wavelength degrees of freedom,
in which case their evolution can be approximated as quasi-instantaneous,
i.e.~quasi-local in time.
However, in the context of LSS both short and large scales evolve on 
the same characteristic timescale $\mathcal{H}^{-1}$. Nevertheless, in perturbation theory the fields in the right hand side of Eq.~\eqref{eq:taunonloc}
can be Taylor-expanded around the fluid trajectory
such that the theory can be reformulated in terms of local-in-time
operators. 
Thus, the effective stress tensor at next-to-leading order is given by~\cite{Carrasco:2012cv,Carrasco:2013mua,Baldauf:2014qfa,Angulo:2014tfa,Bertolini:2016bmt} 
\be 
\label{eq:tauij}
-\frac{1}{a\rho}\d_j\tau^{ij}=-c_s^2 \d^i \delta + \frac{c_v^2}{\mathcal{H}}\d^i\d_kv^k
-c_1\d^i \delta^2 -c_2\d^i (s^{kl}s_{kl}) -c_3 s^{ij}\d_j \delta
-\frac{1}{a\rho}\d_j\tau^{ij}_{\text{stoch.}}\,,
\ee
where $c_s^2,c_v^2$, $c_{1,2,3}$ are time-dependent Wilson coefficients,
and we have introduced the tidal tensor as 
\be
s_{ij}=\frac{2}{3\Omega_m \mathcal{H}^2}\left(\d_i \d_j \Phi -\frac{1}{3}\delta_{ij}\Delta \Phi \right)\,.
\ee
The general basis of counterterms at higher orders involves 
convective derivatives~\cite{Baldauf:2014qfa,Mirbabayi:2014zca,Abolhasani:2015mra}, which come from expanding $\xvec_{\rm fl}$ in \eqn{eq:taunonloc}.
$\tau^{ij}_{\text{stoch.}}$ is the stochastic contribution which is uncorrelated
with $\delta$. It is local and analytic in space and obeys 
the equivalence principle, as well as the 
mass and momentum conservation. At the lowest order it is given by
\be
\label{eq:DM_stochastic}
\d_i\left[\frac{1}{a\rho} \d_j \tau^{ij}_{\text{stoch.}} \right]= J_0 \,,\quad \langle J_0 (\k)J_0(\k')\rangle \propto  \left(\frac{k}{k_{\rm NL}}\right)^4\,. 
\ee

\subsection{Loop expansion}

\begin{figure}[t]
\centering
\includegraphics[width=0.99\textwidth]{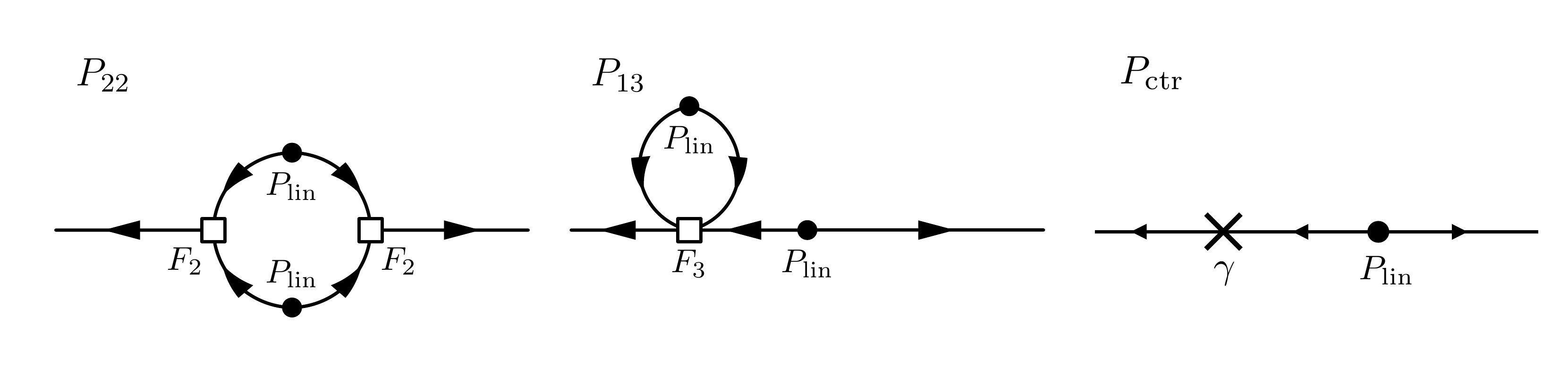}
\caption{\footnotesize Diagrams contributing to the 
deterministic part of the one-loop matter power spectrum.
Taken from Ref.~\cite{Baldauf:2015aha}.}
\label{fig:loopdiag}
\end{figure}

Plugging \eqref{eq:tauij} into \eqref{eq:fluid} we obtain effective 
equations of motion of the matter fluid. At linear order in $\delta, v^i$
it solved by the linear growing mode {}
\be 
\delta_{(1)}= -(\mathcal{H}f)^{-1}\d_i v^i_{(1)}=D(\tau)\delta_0(\k)\,,
\ee
where $\delta_0(\k)$ is the initial density field and $D(\tau)$ is the linear growth factor normalized to unity at zero redshift, $f\equiv d\ln D/d\ln a$ is the logarithmic growth factor. The initial conditions for structure formation are set after recombination, such that $\delta_0(\k)$ is 
a nearly Gaussian random field, whose properties are encoded
in the linear power spectrum $P_{\rm lin}$:
\be
\langle \delta_0(\k) \delta_0(\k')\rangle =(2\pi)^3\delta^{(3)}(\k+\k')P_{\rm lin}(k)\,,
\ee
such that at leading order (in linear theory) we have 
\[
\langle \delta_{(1)}(\tau,\k) \delta_{(1)}(\tau,\k')\rangle = (2\pi)^3\delta^{(3)}(\k+\k')D^2(\tau)P_{\rm lin}(k)\,.
\]
To solve Eq.\eqref{eq:fluid} it is convenient 
to work in the EdS approximation, and 
to split the fields of interest $\delta$ and $\theta\equiv -(\mathcal{H}f)^{-1}\d_i v^i$ into two parts.\footnote{In cosmological perturbation theory 
only the longitudinal part of $v^i$ has a growing mode. The transverse 
part decays in linear theory but gets excited at the non-linear level. In principle, it can be taken into account, but its contribution is negligible for most applications~\cite{Mercolli:2013bsa}.
} 
One part is obtained 
upon formally setting the effective stress tensor to zero, while the other 
part will include corrections due to the presence of this tensor. 
This way the total perturbative solution for the matter density can be written as 
\be 
\label{eq:split}
\begin{split}
\delta=[\delta_{(1)}+\delta_{(2)}+\delta_{(3)}+...]+\delta^{\rm c}_{(1)}+...\,,
\end{split}
\ee
and similarly for the velocity divergence $\theta$. The 
$\delta_{(n)}$ corrections are given by
\be
 \delta_{(n)}(\tau,\k)=D^n(\tau)\int_{\q_1...\q_n} F_n(\q_1,...,\q_n)
 \delta^{(3)}\left(\sum_{i=1}^n \q_i-\k\right)\delta_{0}(\q_1)...\delta_{0}(\q_n)\,,
\ee
where $F_n(\q_1,...,\q_n)$ are certain convolution kernels whose form is dictated by the non-linear structure of the pressureless fluid equations. Explicitly for the first three kernels we have:
\be 
\begin{split}
F_1(\q)=1\,,\quad F_2(\q_1,\q_2)=\frac{5}{7}+\frac{(\q_1\cdot \q_2)}{2}\left(\frac{1}{q_1^2}+\frac{1}{q_2^2}\right) + \frac{2}{7}\frac{(\q_1\cdot \q_2)^2}{q_1^2 q_2^2}\,.
\end{split}
\ee
The first correction generated by the stress-tensor is given by 
\be
\delta_{(1)}^{\rm c}=-\gamma k^2 \delta_{(1)}(\tau,\k) \,,\quad \gamma\equiv 
-
\frac{1}{D(a)}
\int da'D(a') G_\delta(a,a')(c_s^2 +c_v^2)\,,
\ee
where $G_\delta(a,a')$ is the density field Green's function 
of the linearized fluid equations~\cite{Carrasco:2012cv}.
The two-point function of the matter field including leading order non-linearities (i.e.~one-loop corrections) is given by 
$P_{\rm NLO}=D^2(\tau)P_{\rm lin}(k)+ P_{\rm 1-loop}(\tau, k)$ with
\be 
\label{eq:1Loop}
\begin{split}
&P_{\rm 1-loop}(\tau,k)=
2D^4(\tau)\int_\q F_2^2(\k-\q,\q)P_{\rm lin}(|\k-\q|)P_{\rm lin}(q)\\
&+6 D^4(\tau)P_{\rm lin}(k)\int_\q F_3(\k,-\q,\q)P_{\rm lin}(q)-2 
\gamma(\tau) k^2 ~D^2(\tau)P_{\rm lin}(k)+c_{\rm stoch} k^4\,.
\end{split}
\ee
This correction admits a representation in terms of Feynman diagrams
shown in Fig.~\ref{fig:loopdiag}.

The split of the perturbative solution \eqref{eq:split}
is useful for the EFT power counting. On mildly-nonlinear scales the linear power spectrum
can be approximated as a power-law $P_{\rm lin}=(k/k_{\rm NL})^n$ with $n\approx -1.5$~\cite{Pajer:2013jj,Mercolli:2013bsa}. Using the approximate Lifshitz symmetry the 
dimensionless power spectrum can be written as,
\be
\begin{split}
\Delta^2(k) = &\left(\frac{k}{k_{\rm NL}}\right)^{n+3}\left(
1+\left(
\frac{k}{k_{\rm NL}}
\right)^{n+3}
\left[a_1+a_2\ln
\frac{k}{k_{\rm NL}}
\right] \right)\\
&+b_c \left(\frac{k}{k_{\rm NL}}\right)^{n+5}
+b_{\rm stoch} \left(\frac{k}{k_{\rm NL}}\right)^{7}+...\,,
\end{split}
\ee
where the first line contains one-loop corrections produced by the intrinsic 
non-linearity of the fluid equations, while the second line displays the 
terms coming from the deterministic and stochastic parts of the effective 
stress tensor. The two-loop corrections scale as 
\be 
\Delta^2_{\rm 2-loop}(k)\sim  \left(\frac{k}{k_{\rm NL}}\right)^{2(n+3)}\,,
\ee
which indeed confirms that at NLO we only need to keep the effective operators
with Wilson coefficients 
$c_s^2$ and 
$c_v^2$. Note, however, that the actual power spectrum
of our Universe is not a power-law. 
In particular, it has the BAO wiggles, which break the naive 
power counting in $k/k_{\rm NL}$
and require a special treatment within 
a procedure called IR resummation.

\subsection{UV renormalization and IR resummation} \label{uvirsec}

The UV limit of the one-loop integral in Eq.~\eqref{eq:1Loop} reads 
\be
P_{\rm 1-loop}(\tau , k) \Big|_{\rm UV}=-\frac{61 }{630\pi^2} D^4 ( \tau )  k^2P_{\rm lin}(k)
\int_{k \ll q} q^2 dq\,\frac{P_{\rm lin}(q)}{q^2}\,.
\ee
At face value, UV modes couple to modes with mildly-nonlinear wavenumbers
$k
\sim 0.1~h$/Mpc
through the variance of the short mode displacement field.
We see that this integral diverges for a generic initial power spectrum. This divergence is exactly canceled by the Wilson coefficient $\gamma(\tau)$, which ensures that
the physically observed quantities such as the density field n-point 
correlation functions are finite. Their
dependence on short-scale physics is captured by the finite part 
of $\gamma(\tau)$, which can been accurately measured in N-body simulations~\cite{Carrasco:2012cv,Foreman:2015lca,Baldauf:2015aha,Chudaykin:2020hbf} or can be inferred from the data.

The IR limit of the one-loop integral reads~\cite{Baldauf:2015xfa,Blas:2016sfa}:
\be
P_{\rm 1-loop} ( \tau , k ) \Big|_{\rm IR}=
D^4(\tau ) \int_{\q:~q\ll k}   P_{\rm lin}(q)\frac{(\k\cdot \q)^2}{q^4}\left(e^{-\q\cdot \nabla_{k'}}-1\right)P_{\rm lin}(k')\Bigg|_{\k'=\k}\,.
\ee
If the linear power spectrum did not have any feature i.e.~$P_{\rm lin}=P_{\rm smooth}$, such that $\d_k P_{\rm smooth}(k)\sim (1/k)P_{\rm smooth}(k)$, the differential operator above could be Taylor-expanded and we would find that the IR modes couple to a mildly-nonlinear mode $k$ though the 
variance of the large-scale density field~\cite{Jain:1995kx,Scoccimarro:1995if,Blas:2013bpa,Carrasco:2013sva,Creminelli:2013poa}, 
\be 
P_{\rm 1-loop,~smooth} ( k ) \Big|_{\rm IR}\sim P_{\rm smooth}(k)\int_{q\ll k}q^2dq\, P_{\rm smooth}(q)\,.
\ee
This coupling is rather weak. However, $P_{\rm lin}$ contains BAO wiggles, 
whose coupling to IR modes is enhanced. Approximating 
$P_{\rm lin}=P_{\rm smooth}+P_{\rm wiggly}$ with
\mbox{$P_{\rm wiggly}\propto \cos(kr_{\rm BAO})$}, ($r_{\rm BAO}\simeq 110$ $h^{-1}$Mpc is the comoving acoustic horizon at decoupling) we obtain
\begin{align}
P_{\rm 1-loop,~wiggly} ( \tau , k) \Big|_{\rm IR}
&= -\Sigma^2 k^2 D^2(\tau) P_{\rm wiggly} ( k )  \label{eq:1lwiggly}
\\
 & \hspace{-1.1in} \equiv
-\left[ \frac{D^2(\tau)}{6\pi^2}\int_{q\ll k}dq P_{\rm lin}(q)
\left(1-j_0(qr_{\rm BAO})+2j_2(qr_{\rm BAO})\right) \right]
\, k^2 D^2(\tau) P_{\rm wiggly}(k)\,. \nonumber
\end{align}
The integral $\Sigma^2$ 
receives contributions from modes all the way up to $k$,
and it is numerically close to 
the large-scale variance of the displacement field, 
which turns out to be quite large, i.e.~$k^2\Sigma^2\sim \mathcal{O}(1)$ at $z\sim 0$
for modes of interest $k\sim 0.1~h$/Mpc.
Hence, the higher order soft loop corrections to \eqref{eq:1lwiggly}
are not negligible and must be resummed for the correct description of the BAO. This procedure is called ``IR resummation''~\cite{Senatore:2014via,Baldauf:2015xfa,Blas:2016sfa,Vlah:2015zda,Perko:2016puo,Senatore:2017pbn,Ivanov:2018gjr,Lewandowski:2018ywf}. 
It was originally formulated 
within the Lagrangian effective field theory, 
but shortly it was shown that IR resummation can be performed directly at the diagrammatic level within the Eulerian EFT~\cite{Blas:2016sfa,Ivanov:2018gjr}. At zeroth order in hard loops (with $q\gtrsim k$) one has 
\be 
P_{\rm IR-res,~wiggly}(\tau,k)=e^{-k^2\Sigma^2}(D^2 ( \tau)  P_{\rm wiggly}(k))\,.
\ee

\subsection{Flavors of the EFTs}

It is important to stress that at the technical level, 
there are several different ways to realize the EFT of LSS ideas.
The original proposal of the EFT in Eulerian 
fluid variables that we have discussed so far 
is plagued by the large IR contributions
that require IR resummation. 
IR resummation in terms of Eulerian fluid variables 
is complicated by the presence of the 
spurious IR enhancements in the loop diagrams.
This motivated the development 
of the Lagrangian EFT of LSS~\cite{Porto:2013qua,Vlah:2015sea,Vlah:2015zda,Vlah:2016bcl,Vlah:2018ygt,Chen:2020zjt,Chen:2020fxs}.
The Lagrangian EFT of LSS also partially resums 
some of the UV contributions. 
From the computation efficiency point of view, however, it is still 
beneficial to work in Eulerian space. 
In this case it is still possible to perform a systematic 
IR resummation, as discussed in \secref{uvirsec}, which is particularly manifest within a path integral formulation
of the EFT of LSS known as Time-Sliced Perturbation Theory~\cite{Blas:2015qsi,Blas:2016sfa,Ivanov:2018gjr,Vasudevan:2019ewf}.
All these different techniques agree within
the overlapping domains.
This reflects the uniqueness
property of the EFT: the predictions for physical 
processes do not depend on a particular formulation,
once the results are compared to the same order
in appropriate small parameters. In other words, at a given order 
in relevant IR and UV small parameters, the difference between the 
EFT formulations appears only at higher orders.

\subsection{Biased tracers}

So far we have discussed the 
clustering of pure matter. 
The galaxy density field observed in 
cosmological surveys is a biased tracer 
of the underlying dark matter field.
The relationship between them is given by a 
past light-cone integral over local long wavelength perturbations of 
matter density, velocity, 
and tidal fields~\cite{Senatore:2014eva,Mirbabayi:2014zca,Assassi:2014fva,Angulo:2015eqa,Desjacques:2016bnm,Nadler:2017qto},
\be
\label{eq:deltag1}
\begin{split}
\delta_g=&\int^\tau d\tau'
\mathcal{H}(\tau')
\Big[
\beta_1(\tau,\tau')
\delta(\tau', \x_{\rm fl})
+\beta_2(\tau,\tau')\mathcal{H}^{-1}\d_iv^i(\tau',\x_{\rm fl})
+\beta_3(\tau,\tau')\delta^2(\tau',\x_{\rm fl})\\
&
+\beta_4(\tau,\tau')R_*^2\d^2_{\x_{\rm fl}}\delta(\tau',\x_{\rm fl})
+\beta_5(\tau,\tau') \mathcal{H}^{-4}(\d_{i}\d_j\phi)^2(\tau',\x_{\rm fl})+
\beta_6(\tau,\tau')\epsilon(\tau',\x_{\rm fl})+...
\Big]\,,
\end{split} 
\ee
 where $\epsilon$ is a random field uncorrelated with $\delta$, which 
captures the stochasticity of the tracer, $\beta_i$ are time-dependent kernels with characteristic 
timescale $\mathcal{H}^{-1}$ and order-one amplitudes, i.e.~$\d_\tau \beta_i\sim \mathcal{H}$, $R_*$ is the typical length scale of the object.
Just like in the case of the effective stress-tensor of matter that we discussed above,
the apparent non-locality in time in
Eq.~\eqref{eq:deltag1} can be removed in perturbation theory
by Taylor-expanding around the fluid trajectory, 
which allows one to rewrite the bias relation as a local-in-time
expression~\cite{Assassi:2014fva}\footnote{Strictly speaking, it is possible to rewrite the bias
expansion in the local-in-time form only in the so-called Einstein-de-Sitter approximation for the time evolution~\cite{Desjacques:2016bnm}, which is $\mathcal{O}(0.1\div 1)\%$ 
accurate for redshifts relevant to current and future surveys.  For works dealing with exact time dependence, see for example \cite{Fasiello:2016qpn,Lewandowski:2017kes,Donath:2020abv}. }
\be 
\begin{split}
\delta_g=b_1 \delta + \frac{b_2}{2}\delta^2 +b_{\mathcal{G}_2}\mathcal{G}_2+b_{\nabla^2\delta }\nabla^2\delta+b_{\Gamma_3}\Gamma_3+\varepsilon+...\,,
\end{split} 
\ee
where $\varepsilon$ is the stochastic field uncorrelated with the long-scale perturbations, 
and we have introduced new Galileon operators
\be 
\begin{split}
& \mathcal G_2 (\Phi)\equiv (\d_i \d_j \Phi)^2 - (\Delta\Phi)^2\,,\quad \Gamma_3\equiv \mathcal G_2 (\Phi)-\mathcal G_2 (\Phi_v)\,,
\end{split} 
\ee
($\Phi_v$ is the velocity potential),
and ``$...$'' denote both the operators which do not contribute  
to the one-loop power spectrum after renormalization and higher order operators.
Free parameters $b_1,b_2, b_{\mathcal{G}_2},b_{\nabla^2\delta},b_{\Gamma_3}$
are time-dependent 
Wilson coefficients. In the cases where the bias tracers are galaxies or dark matter halos,
these bias parameters, up to quartic order,
have already been detected 
in simulations, see e.g.~\cite{Lazeyras:2015lgp,Lazeyras:2017hxw,Desjacques:2016bnm,Abidi:2018eyd,Lazeyras:2019dcx,Nishimichi:2020tvu}.

Importantly, the bias expansion has a new UV scale $R_*$, which can be associated with the typical 
size of the collapsed object~\cite{Mirbabayi:2014zca,Senatore:2014eva,Desjacques:2016bnm,Lazeyras:2019dcx}. 
For halos this has the order of magnitude of the
Lagrangian radius of the overdensity clump that collapses into a host halo.
For the line emission this has the order of the Jeans scale of the diffuse gas~\cite{Desjacques:2016bnm,Cabass:2018hum}.
For galaxies $R_*$ depends both on the host halo properties and 
on the details of galaxy formation, e.g.~the ambient radiation field and thermal heating of intergalactic medium.

Another important difference compared to the dark matter case is that the power spectrum of the stochastic field~$\epsilon$ does not fall off on large scales as in Eq.~\eqref{eq:DM_stochastic}, but it is rather constant as $k$ goes to zero. This is a consequence of the fact that for biased tracers mass and momentum are not conserved, given that each galaxy is counted the same, regardless of the mass of the host dark matter halo. This constant power spectrum of the stochastic field is related to the well known Poisson noise for biased tracers.

\subsection{Redshift-space distortions}

When galaxy surveys map the Universe they assign the radial position of galaxies according to their redshifts. 
The observed galaxy redshift is contaminated by peculiar 
velocity projections onto the line-of-sight, which gives rise to redshift-space distortions (RSD).
From the EFT point of view, RSD boil down to the following velocity-dependent 
coordinate transformation
\be
\begin{split}
\delta^{(z)}(\tau,\k)=\delta(\tau,\k)
+\int d^3x~ e^{-i\k\cdot \x}\left(
\exp\left[-i\frac{k_z}{\mathcal{H}}v_z(\tau,\x)\right]-1
\right)(1+\delta(\tau,\x))\,,
\end{split}
\ee
where $z$ denotes the line-of-sight direction
and we have employed the 
plane-parallel approximation valid on short scales. Taylor expanding the exponent in the RSD mapping 
and coarse graining the resulting composite operators involving 
various insertions of velocity fields one obtains a set of new Wilson coefficients 
which properly renormalize the UV sensitivity
of the redshift-space density.
Thus, the redshift-space mapping is an additional source 
of non-linearity which can be consistently taken into account within 
the EFT~\cite{Senatore:2014vja,Lewandowski:2015ziq,Perko:2016puo}.

\subsection{Baryons in the EFT of LSS}

One particularly compelling advantage of the EFT framework for LSS is that it is possible to include analytically the effects of small scale baryonic, or star-formation, physics on large-scale clustering \cite{Lewandowski:2014rca,Braganca:2020nhv}.  In this approach, one treats the CDM and baryons as separate fluids coupled through gravity, each with its own set of EFT parameters which capture the UV properties of the system.   The functional form of these effects on large scales, i.e.~as a function of $k$, is fixed by symmetries and organized in a controlled derivative expansion, just like the the pure CDM case described above.  For predictions on large scales, this can be a significant advantage over relying on $N$-body simulations that include baryonic processes.  This is because, unlike the case for pure CDM, we do not know a priori the short-scale baryonic physics that should be included in the simulations.

\begin{figure}[t]
\centering
\begin{tabular}{cc}
\hspace{-.4in}\includegraphics[width=16cm]{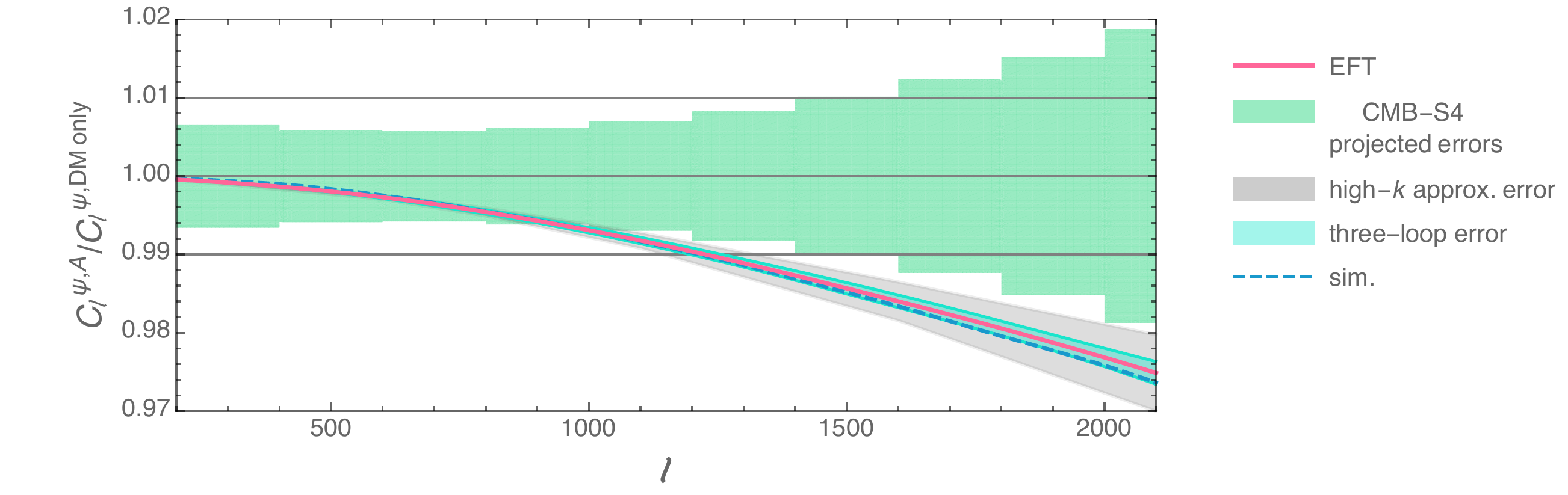} 
\end{tabular} 
\caption{\footnotesize  From \cite{Braganca:2020nhv}.  \emph{Caption edited for context:} 
Ratio of the total baryon-plus-dark-matter and dark-matter-only lensing potential power spectra $C_{\ell}^{\psi, A}/ C_{\ell}^{\psi, \text{DM only}}$ using the two-loop EFT predictions.
 {The green band is the projected error for CMB-S4 \cite{Abazajian:2016yjj}, the gray band is the estimated theory error coming from the `high-$k$ approximation' described in \cite{Braganca:2020nhv}, and the teal band is the estimated error coming from the three-loop terms in the EFT.}  The dashed blue line is the result of direct numerical integration of the outputs of the simulations.  We see that CMB-S4 will be highly sensitive to the effects of baryons on the lensing potential, and that the two-loop EFT can reliably capture these effects up to $\ell \lesssim 2000$, and actually even beyond. }
\label{lensing3}
\end{figure}

As an example, \cite{Braganca:2020nhv} used this approach to compute the CMB lensing potential and compare with a numerical simulation (see \figref{lensing3}), which shows quite good agreement up to $\ell \approx 2000$.  This is an important observable because it can be used to probe neutrino masses in CMB-S4, and much of the constraining power for a mass sum of less than $120 \text{ MeV}$ comes from $\ell \lesssim 2100$ \cite{Abazajian:2016yjj}.

\subsection{EFT of LSS at the field level}

So far we have been focusing on calculation of correlation functions. However, the EFT of LSS naturally predicts the full non-linear density field, given some realization of the initial conditions. This can be exploited in two ways. 

First, the field-level predictions provide a natural way to compare the theory to numerical simulations. If they share the same initial conditions, the comparison can be done without paying the price of cosmic variance. Furthermore, such comparison is much more stringent since one has to fit all Fourier modes and not only the summary statistics. This has been exploited in the past to provide the first reliable measurements of the EFT parameters and the nonlinear scale~\cite{Baldauf:2015tla,Baldauf:2015zga} as well as for the detailed comparison of theory and simulations for biased tracers in real and redshift space~\cite{Schmittfull:2018yuk,Schmittfull:2020trd}. 

Second, these methods can be used to construct the field-level 
EFT likelihood in the perturbative forward modelling. Such approach aims at measuring cosmological parameters from the full nonlinear field, without using summary statistics. A lot of progress has been made recently towards achieving this goal, see for instance~\cite{Schmidt:2018bkr,Elsner:2019rql,Cabass:2019lqx,Schmidt:2020ovm,Cabass:2020nwf,Schmidt:2020viy,Schmidt:2020tao,Nguyen:2020hxe,Cabass:2020jqo}.

Finally, some progress was made recently in fixing the form of galaxy correlation functions using only symmetries of the system and the equivalence principle, without explicitly relying on the equations of motion. This is inspired by the similar cosmological bootstrap approach to derive the form of inflationary correlators from symmetries and general principles such as locality and unitarity. The natural starting point for this ``LSS bootstrap'' is at the field level, where various theoretical constraints can be straightforwardly imposed~\cite{Fujita:2020xtd,DAmico:2021rdb}.

\subsection{Extensions}

Other important extensions of the EFT include the incorporation of 
non-Gaussian initial conditions~\cite{Lewandowski:2015ziq,Angulo:2015eqa,Assassi:2015jqa,Assassi:2015fma},
IR-resummation of primordial oscillating features~\cite{Vasudevan:2019ewf,Beutler:2019ojk,Chen:2020ckc},
and an accurate treatment of massive neutrinos.
The later is a conceptually challenging task, as the
neutrino free-streaming scale $l_{\rm fs}$ is significantly longer than 
the non-linear scale $k_{\rm NL}^{-1}$.
However, massive neutrinos can be split into ``fast''
and ``slow'' ones, which allows to identify a small
parameter in the regime $k>l_{\rm fs}^{-1}$ and systematically compute 
their effect 
on dark matter clustering~\cite{Senatore:2017hyk,deBelsunce:2018xtd}.

Another important task is to account for selection 
effects, which may be present in realistic surveys.
The effective operators capturing these effects at leading 
orders are given in Ref.~\cite{Desjacques:2018pfv}.
The extensions for CMB lensing, galaxy lensing, and intrinsic 
alignments are worked out in Refs.~\cite{Foreman:2015uva,Vlah:2019byq}.  The incorporation of additional degrees of freedom associated 
with dark energy and modified gravity was done in
Refs.~\cite{Lewandowski:2016yce,Cusin:2017mzw,Cusin:2017wjg,Bose:2018orj,Crisostomi:2019vhj,Lewandowski:2019txi} and is described in more detail in \secref{eftdesec}.

\subsection{State-of-the-art computations}

The state-of-the-art EFT calculations that have been carried out up to now 
are listed in Table~\ref{tab:calc}, see Refs.~\cite{Carrasco:2013mua,Carrasco:2013sva,Foreman:2015lca,Bertolini:2015fya,Bertolini:2016bmt,Eggemeier:2018qae,Konstandin:2019bay,Steele:2020tak,Steele:2021lnz,Ivanov:2021kcd,Baldauf:2021zlt}. These calculations must be
extended to higher loop orders and higher $n$-point functions
in order to use more observed modes in cosmological data analyses.
The consistent inclusion of massive neutrinos has been
done for the one-loop power spectrum and tree-level matter bispectum~\cite{Senatore:2017hyk,deBelsunce:2018xtd}.
The results of IR resummation formally exist for an arbitrary n-point function
and for an arbitrary number of hard loops both in real and redshift spaces
and for a generic biased tracer~\cite{Ivanov:2018gjr}.

\begin{table}[t]
\begin{center}
\begin{tabular}{|c|c|c|c|} 
 \hline
Type & Power spectrum & Bispectrum  & Trispectrum  \\ [0.5ex] 
 \hline\hline
 Matter in real space & 3-loop & 2-loop & 1-loop \\ \hline
  Biased tracers in real space & 1-loop & 1-loop & --- \\ \hline
  Biased tracers in redshift space & 1-loop & 0-loop & ---\\ \hline
\end{tabular}
\caption{ \footnotesize Available EFT calculations  
for the two-, three-, and four-point functions of various tracers.
}
 \label{tab:calc}
\end{center}
\end{table}
It is worth mentioning that the computation of EFT loop corrections
requires efficient numerical tools to evaluate perturbation theory 
convolution integrals.
These techniques are
necessary in order to apply the EFT calculations 
to observational data. FFTLog is one of such techniques~\cite{McEwen:2016fjn,Fang:2016wcf,Simonovic:2017mhp}.
It has been worked out to one-loop order for biased tracers in redshift space~\cite{Chudaykin:2020aoj}, for the two-loop power spectrum and the one-loop
bispectrum for matter in real space~\cite{Simonovic:2017mhp}, and for BAO resummation in \cite{Lewandowski:2018ywf}.

Finally, combining all these efforts, a several independent codes have been written with the aim at providing efficient and reliable state-of-the-art EFT computations~\cite{Chudaykin:2020aoj,DAmico:2020kxu,Chen:2021wdi}.

\subsection{Applications to current and future data} \label{datasec}

The EFT of LSS allows one to take advantage of the 
cosmological information encoded in the full shape of the observed
galaxy power spectrum. This means
a consistent analysis of the large-scale structure data 
that includes fitting fundamental cosmological parameters 
directly from the power spectrum shape,
as it is routinely done in cosmological analyses of the CMB.\footnote{This can be contrasted with the nomenclature of 
some previous works, which used the term ``full shape''
for an analysis which studies how a particular fixed power spectrum  
template gets distorted by the Alcock-Paczynski effect~\cite{Alcock:1979mp}.}
This analysis has been done for the first time in Refs.~\cite{Ivanov:2019pdj,DAmico:2019fhj},
which showed that galaxy power spectrum measurements from the Baryon
acoustic Oscillation Spectroscopic Survey (BOSS) data release 12~\cite{Alam:2016hwk} is a powerful
source of cosmological information.

\begin{figure}[t]
\centering
\includegraphics[width=0.75\textwidth]{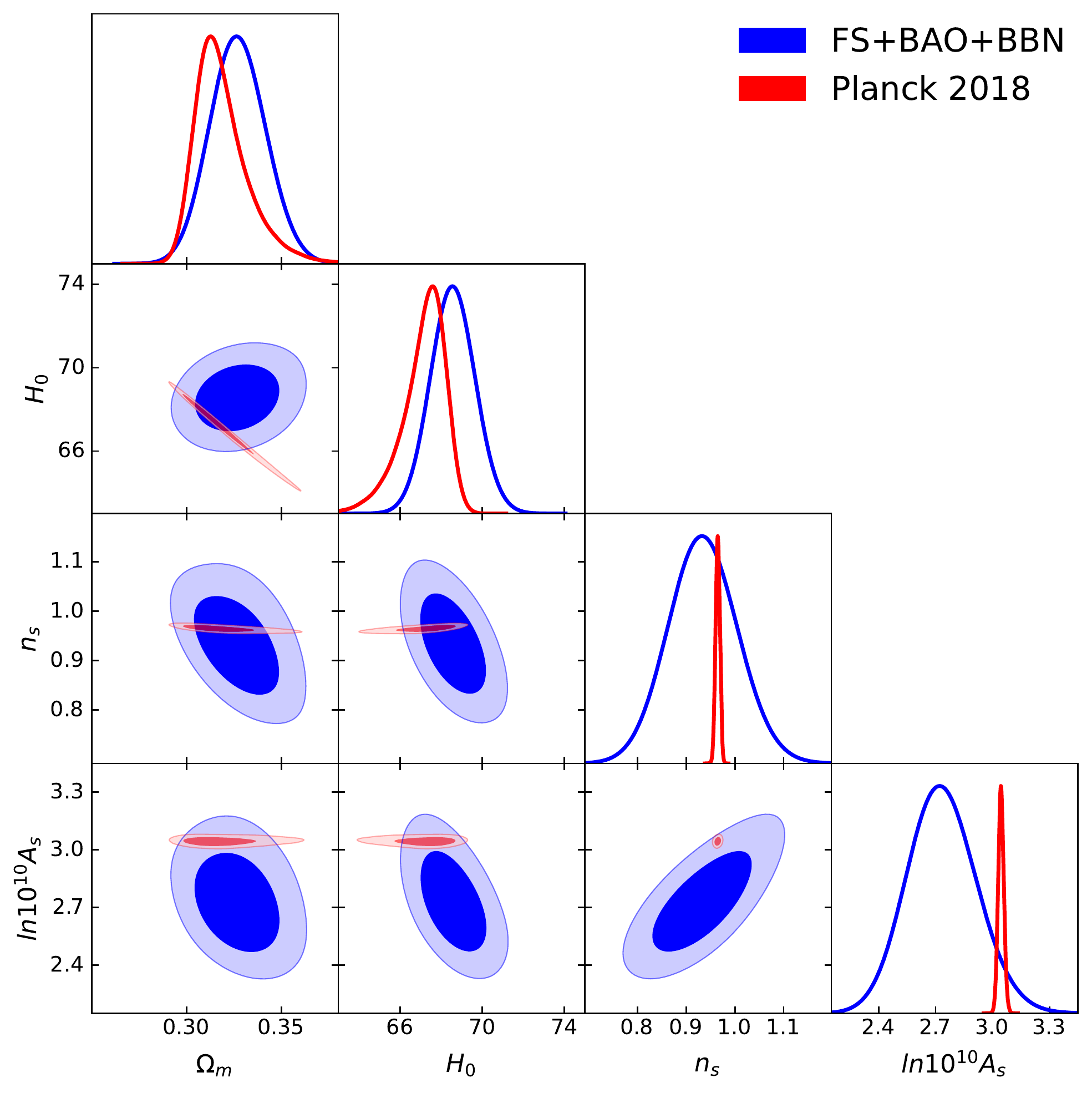}
\caption{\footnotesize Cosmological parameters of the base $\Lambda$CDM model 
as measured from the EFT-based full-shape (FS) BOSS DR12 and eBOSS ELG 
redshift-space galaxy power spectrum likelihoods combined with final 
post-reconstruction BAO 
measurements from BOSS and eBOSS surveys and the BBN baryon density
prior (FS+BAO+BBN, in blue). 
The results of the Planck CMB 2018 baseline analysis~\cite{Aghanim:2018eyx} are shown for comparison
in red. See Ref.~\cite{Ivanov:2021zmi} for more detail.
}
\label{fig:params}
\end{figure}

EFT-based analyses of the BOSS data yield the CMB-independent
measurements of the parameters of the base $\Lambda$CDM model and its extensions~\cite{Aghanim:2018eyx}:
the Hubble constant $H_0$, the current matter density fraction 
$\Omega_m$, the primordial power spectrum amplitude $A_s$ and tilt $n_s$,
the mass fluctuation amplitude $\sigma_8$,  as well as constraints on
the spatial curvature of the Universe $\Omega_k$, and
the dark energy equation of state parameters~\cite{Philcox:2020vvt,Philcox:2020xbv,Colas:2019ret,Chudaykin:2020ghx,DAmico:2020kxu,DAmico:2020tty,Ivanov:2021haa,Ivanov:2021zmi,Philcox:2021kcw,Chen:2021wdi}. 
Remarkably, many of these parameters are measured with precision similar 
to that of the Planck
CMB data results, e.g.~$H_0$ and $\Omega_m$, see Fig.~\ref{fig:params}. 
Besides, combining the EFT-based full-shape BOSS likelihood 
with the CMB data has lead to new constraints on the total neutrino mass,
effective number of relativistic degrees of freedom~\cite{Ivanov:2019hqk}.
Moreover, the new BOSS likelihood allowed one to derive new constrains
on certain models addressing the so-called ``Hubble tension.'' The use of the EFT-based likelihood was crucial in order to show that such models are ruled out by the current 
large-scale structure data~\cite{Ivanov:2020ril,DAmico:2020ods}.

 In addition, another exciting development has been the application of the EFT of LSS to constrain primordial non-Gaussianities (see \secref{subsec:PNG_shapes})~\cite{Cabass:2022wjy,DAmico:2022gki},
including the first-ever bounds 
on single-field primordial non-Gaussianity
from galaxy surveys.  Local-type non-Gaussianity, typical for multifield models, has been constrained using the scale-dependent bias of the power spectrum in, for example,~\cite{Slosar:2008hx,Xia:2011hj,Castorina:2019wmr}.

The sensitivity forecasts for ongoing experiments such as Euclid and DESI 
suggest that the application of the EFT to these surveys can lead to 
significant improvements in cosmological parameter measurements~\cite{Chudaykin:2019ock}.
This includes a $5\sigma$-detection of the sum of neutrino masses
and $0.1\%$ measurement of the Hubble constant from the combination
of the Planck CMB and Euclid/DESI data.  Also, recent constraints on non-Gaussianity suggest promising and competitive results in the future, given the volume of ongoing state-of-the-art surveys and the ability of EFT formalism to provide more precise computation of relevant observables. These conclusions are based 
on a realistic analysis including marginalization over all 
necessary Wilson coefficients and data cuts consistent with the
theoretical error~\cite{Baldauf:2016sjb},
which is determined by calculations that are 
available at present. The results are expected to 
improve with more precise calculations and with better priors
on Wilson coefficients, which can be obtained from high fidelity
numerical simulations~\cite{Wadekar:2020hax}. Indeed, these types of analyses, and their higher-precision versions of the future, were one of the main motivations for the development of the EFT of LSS.  

Looking beyond this decade to the next generation of high-redshift spectroscopic surveys, an increase by another order of magnitude in the number of observed galaxies is expected (for example, see the snowmass white paper on opportunities of high-redshift and large-volume future surveys~\cite{Snowmass2021:highz} and references therein). In this coming era of the ultimate precision, the EFT methods discussed here will be even more valuable.

\section{Effective Field Theory of Dark Energy}\label{eftdesec}

\noindent  Observationally speaking, Einstein's theory of General Relativity (GR), as far as we can tell, successfully describes gravitational and cosmological phenomena over an enormous range of length and time scales.  For example, GR describes small effects in our Solar System, such as the precession of the perihelion of Mercury and the bending of light around the Sun, it describes large effects like the expansion of the Universe, both at early and late times, it describes gravitational-wave emission by binary inspirals, and it describes black holes.   The broad applicability of GR, however, is (most likely) not simply a convenient accident.  GR is the unique low-energy Lorentz invariant theory of an interacting massless spin-2 particle (see e.g.~\cite{PhysRev.138.B988,weinberggravandcosmo}), and gauge invariance of the action implies that all other fields couple to gravity with the same strength (this is called the equivalence principle, see e.g.~\cite{PhysRev.135.B1049}).  These facts make the predictions of the universal, long-range force quite robust.

The standard cosmological paradigm, describing the large-scale evolution of the Universe from the moments after the Big Bang until the current time, is GR with a cosmological constant $\Lambda$ coupled to a fluid-like system of cold dark matter (CDM) particles, called $\Lambda$CDM.   This model so far successfully describes cosmological phenomena such as the cosmic microwave background, Big Bang nucleosynthesis, the large-scale structure of the Universe, and gravitational lensing of galaxies, to name a few.

A historically theoretically worrying critique of the \lcdm paradigm, though, is the cosmological constant problem.  This is the fact that the observed value of the background energy density $\Lambda^4$ is 60 to 120 orders of magnitude smaller than what is expected from our understanding of particle physics, and seems to represent a huge fine-tuning problem for the theory \cite{RevModPhys.61.1}.  In response to this problem, Weinberg suggested the compelling anthropic solution  \cite{PhysRevLett.59.2607,Weinberg:2000yb}.  His argument roughly goes like this.  First of all, viewing GR as a low-energy effective field theory, the cosmological constant $\Lambda$ is the most relevant operator, and so is generically expected to be present in the low-energy theory, although its value is not known \emph{a priori}.  In order to estimate its value, Weinberg pointed out that if it were much larger than the current observed value, there would be no stars or planets (and therefore no humans) in the Universe.   Thus, if there are many patches of the Universe with different values of $\Lambda$ (for example, in a multiverse scenario), then humans would only exist in the patches with small values of $\Lambda$.

This then brings us to dark energy (DE) and modified gravity theories,\footnote{For the purposes of this review, we do not make a meaningful distinction between DE and modified gravity theories.} which are essentially attempts to change the low-energy dynamics of gravitation.  Historically speaking, these theories were considered, in part, as potential explanations for various perceived theoretical shortcomings of GR, such as in the Brans-Dicke theory \cite{Brans:1961sx} and early quintessence models \cite{Fujii:1982ms}.  Theories of DE and modified gravity attempt, among other things, to solve the cosmological constant problem\footnote{Although no compelling solution has yet been found.} or explain the acceleration of the Universe without an explicit cosmological constant (so-called self-accelerating solutions, see for example \cite{Chow:2009fm}).  While the history of motivations to study extensions to GR (see \cite{Joyce:2014kja} for example for a review) is important, we prefer to take a slightly different perspective in this review, one more in line with the EFT principles that we have been discussing.  Here we simply ask \emph{what are the possible observable deviations from $\Lambda$CDM?}  The question in this form has the advantage that it points us toward systematic ways in which we can test GR and look for deviations caused by new physics, which is especially relevant in this era of precision cosmological measurements.\footnote{Upcoming observations range from galaxy surveys like the Rubin Observatory (formally LSST), Euclid, and DESI, to CMB measurements with CMB-S4, to 21cm emission measurements with SKA, to measurements of gravitational waves with LIGO/Virgo, together representing billions of dollars of international investment.}

%
\subsection{Unitary-gauge action in the presence of matter}

In this review, we focus on modifications to \lcdm arising from an extra scalar mode which is related to the breaking of time diffeomorphisms in the Universe (i.e.~the presence of the preferred slicing of space-time where the CMB is nearly homogenous and isotropic).  As discussed above in \secref{subsec:unitary_gauge_action}, a general way of describing all such possible modifications is to write the action for the metric in unitary gauge.  Instead of demanding that the action be diffeomorphism invariant, we demand that it be invariant under time-dependent spatial diffeomorphisms.  Since the action has less symmetry, there is an extra scalar degree of freedom in addition to the normal two degrees of freedom of the graviton, and the scalar mode can be made manifest by performing the Stueckelberg trick.  In fact, this procedure is exactly the same as the one discussed in \secref{subsec:unitary_gauge_action} for the EFTI, with the only difference now being that, because we want to describe late-Universe physics, we have to include the coupling of the metric to matter.  

The full action $S$ is made up of a gravitational part $S_G$, and a matter part $S_M$, 
\be \label{totalaction}
S = S_G + S_M \ .
\ee
In terms of covariant objects like the Riemann tensor $R_{\mu \nu \rho \sigma}$ and the covariant derivative $\nabla_\mu$, as well as time-diffeomorphism breaking operators like $g^{00}$ and the extrinsic curvature $K_{\mu \nu}$, the  gravitational action has the form \cite{Creminelli:2006xe, Cheung:2007st, Gubitosi:2012hu}
\be \label{sgexamp}
S_G = \int d^4 x \sqrt{-g} \,  F_G \left( R_{\mu \nu \rho \sigma} , g^{00} , K_{\mu \nu} , \nabla_\mu ; t \right) \ . 
\ee
Here, the index `$0$' referrs to the time coordinate $t$ which parameterizes equal-time surfaces.
The matter action can also in principle depend on all of the aforementioned fields and the matter fields, $\chi_a$, coupled in such a way that allows operators which break time diffeomorphisms.  Thus, the generic form is  (see \cite{Senatore:2010wk} for example)
\be
S_M = \int d^4 x \sqrt{-g} \, F_M \left(R_{\mu \nu \rho \sigma} , g^{00} , K_{\mu \nu} , \nabla_\mu , \chi_a ; t \right), \ 
\ee
with the same rule that for any covariant object, it is allowed to appear with an upper $0$ index.  Once the action is written in this way, the Stueckelberg trick can be used to introduce the scalar mode $\pi$ just as in \eqn{eq:g00_transformation}, for example.

In this review, we assume the existence of a frame, called the Jordan frame, where each matter species is minimally coupled to the same metric.  Then, the action in the Jordan frame in unitary gauge reads 
\be \label{generalaction}
S = S_{G}[g_{\mu\nu}]+ S_M [g_{\mu \nu} , \chi_a],
\ee
where $S_G$ is as in \eqn{sgexamp}, but $S_M$ is fully diffeomorphism invariant. For the matter action, we can write
\be \label{matteraction1}
S_M = -\half \int d^4 x \sqrt{-g} \, T^{(\rm m)}_{\mu \nu} \delta g^{\mu \nu} 
\ee
where for pressureless CDM, we have
\be
T^{(\rm m)}_{\mu \nu} = \rho_{\rm m} u_\mu u_\nu
\ee
where $\rho_{\rm m}$ is the energy density in the rest frame of the fluid, and $u_\mu$ is the fluid four-velocity.  In the non-relativistic limit, we have
\be
T^{(\rm m)}{}^0{}_0 = - \rho_{\rm m} \equiv - \bar \rho_{\rm m} ( 1 + \delta ) \ ,  \quad T^{(\rm m)}{}^0{}_i = \rho_{\rm m} a v^i \ , \quad T^{(\rm m)}{}^i{}_j = \rho_{\rm m} v^i v^j  \ , 
\ee
and we have introduced the background energy density $\bar \rho_{\rm m} (t)$, the overdensity $\delta$ and the fluid three-velocity $v^i$.  

Similar to \eqn{eq:efti-3}, we can write the gravitational action as 
\be\label{action}
S_G = \int d^4 x \sqrt{-g}\bigg[ \frac{M_*^2}{2} f(t) R - \Lambda(t) - c(t) g^{00} \bigg] + S_{DE}^{(2)}  \,,
\ee
where the explicit operators shown are the only ones that contain linear perturbations, while $S_{DE}^{(2)}$ contains terms that start quadratic in the fields.  Furthermore,  $M_*$ is constant and is related to the effective Planck mass by \eqn{effectiveplanck} below.  The presence of the function $f(t)$ above differs from the inflationary case, where, since there is no matter, one can always eliminate $f(t)$ through a redefinition of the metric.  From \eqn{matteraction1} and \eqn{action}, we can then find the background equations (i.e.~to cancel the tadpole terms) which are given by \cite{Gubitosi:2012hu}\footnote{We assume zero spatial curvature of the background FLRW metric throughout for simplicity.}
\begin{align}
\begin{split} \label{candlambda}
c &  =  M_*^2 f \left(  - \dot H - \half \frac{\ddot f}{f} + \frac{H}{2} \frac{\dot f }{f}  \right) - \half \bar \rho_{\rm m}  \ , \\
\Lambda  & =  M_*^2 f \left( \dot H + 3 H^2 + \half \frac{\ddot f}{f} + \frac{5 H}{2} \frac{\dot f}{f} \right) - \half \bar \rho_{\rm m}  \ ,
\end{split}
\end{align}
which reduce to the inflationary relations \eqn{eq:efti-4} when $M_* = \mpl$, $f =1$, and $ \bar \rho_{\rm m} = 0$.  Once the background equations have been determined, the differences among DE theories is contained in $S^{(2)}_{DE}$.  

At this point, in order to generate the most general DE models, one could simply write all of the possible terms in $S^{(2)}_{DE}$ invariant under time-dependent spatial diffeomorphisms, organized in a derivative expansion, much like in \eqn{eq:efti-3}.  However, many DE models in the literature focus on a specific subset of operators, motivated by the following considerations.  First of all, the scale associated with the observed background expansion is 
\be
\Lambda_2 \equiv (H_0 \mpl )^{1/2}   \sim \frac{1}{10^{-7} \, \text{km}} \ .
\ee
Then, because GR has been stringently tested on Solar System scales, one typically tries to set up a screening mechanism (for example Vainshtein screening \cite{Vainshtein:1972sx, Babichev:2013usa}), so that GR is recovered on smaller scales, but gravity is modified on larger cosmological scales.  The Vainshtein mechanism relies on large nonlinear terms in the action which leads to a second scale $\Lambda_3$ defined by
\be
\Lambda_3 \equiv (H_0^2 \mpl)^{1/3} \sim \frac{1}{10^3 \, \text{km}} \ ,
\ee
(which is roughly the Vainshtein radius for a Planck mass) and leads to Galaxy size screening for the Sun, for example.  Thus, we will consider interactions in the EFT which are suppressed by this much smaller scale $\Lambda_3$.

While promoting these nonlinear interactions, higher-derivative terms can generically appear in the equations of motion, which can lead to unstable Ostrogradski ghosts (see \cite{Woodard:2006nt}, for example).  To get around this, it is common to consider models which explicitly only contain second derivatives in the equations of motion (Horndeski theories \cite{Horndeski:1974wa, Deffayet:2011gz}), or that contain higher derivatives, but have a degenerate kinetic structure so that only one extra, non-ghost, scalar mode propagates (beyond Horndeski, Gleyzes-Langlois-Piazza-Vernizzi (GLPV), and degenerate higher-order scalar-tensor (DHOST) theories \cite{Zumalacarregui:2013pma, Gleyzes:2014dya,Gleyzes:2014qga, Langlois:2015cwa, Crisostomi:2016czh, BenAchour:2016fzp}).  The specific choices of nonlinear terms in these theories is protected from large quantum corrections by a weakly broken galileon invariance \cite{Pirtskhalava:2015nla,Santoni:2018rrx}.  Finally, for everything we discuss in this review, we will be working in the Newtonian limit, where $\partial / H \gg 1$, so we will only consider the leading operators in this limit.

With these considerations in mind, a quite general EFT action for the tensor and scalar modes is (for GLPV theories, see \cite{Creminelli:2017sry} and references therein)
\begin{align}
S^{(2)}_{DE}  =  & \int  \text d^4 x \sqrt{-g}   \bigg[   \frac{m_2^4(t)}{2} (\delta g^{00})^2 - \frac{m_3^3(t)}{2} \, \delta K \delta g^{00}  - m_4^2(t) \delta {\cal K}_2 +\frac{\tilde m_4^2(t)}{2} \,\delta g^{00}\, {}^{(3)}\!\R \nonumber \\ 
&\hspace{.1in}  - \frac{m_5^2(t)}{2}  \delta g^{00} \delta {\cal K}_2  -   \frac{m_6(t)}{3} \delta {\cal K}_3  
 - \tilde m_6 (t) \delta g^{00} \delta {\cal G}_2  -   \frac{m_7(t)}{3} \delta g^{00} \delta {\cal K}_3  \bigg] \;, \label{total_action}
\end{align}
with
\be
\label{deltaKK}
\begin{split}
\delta {\cal K}_2 &\equiv \delta K^2 -  \delta K^\nu{}_\mu \delta K^\mu{}_\nu \;, \qquad \delta {\cal G}_2  \equiv  \delta K^\nu{}_\mu \, {}^{(3)}\!R^\mu{}_\nu -  \delta K \, {}^{(3)}\!R/2 \;, \\
\delta {\cal K}_3  &\equiv \delta K^3 -3 \delta K \delta K^\nu{}_\mu \delta K^\mu{}_\nu + 2 \delta K^\nu{}_\mu \delta K^\mu{}_\rho \delta K^\rho{}_\nu   \; ,
\end{split}
\ee
where ${}^{(3)}\!R^\mu{}_\nu$ is the three-dimensional Ricci tensor of the equal-time hypersurfaces and the time-dependent effective Planck mass is given by
\be \label{effectiveplanck}
M^2 ( t) \equiv M_*^2 f(t) + 2 m_4^2 ( t) \ . 
\ee
 For example, Horndeski theories have $\tilde m_4^2=m_4^2$ and $\tilde m_6=m_6$. 
It is sometimes convenient to express the above EFT parameters as dimensionless parameters that are expected to be $\mathcal{O}(1)$, 
\begin{align}
\begin{split}
\label{EFTaction_masses}
 & \alphaB  \equiv \frac{\MM^2 \dot f - m_3^3 }{2 M^2 H} \;,  \qquad
 \alphaM  \equiv  \frac{ \MM^2 \dot f + 2 (m_4^2)^{\hbox{$\cdot$}}}{M^2 H }\;, \qquad  \alphaT  \equiv - \frac{2 m_4^2}{M^2 } \;,   \\
  & \bone   \equiv \frac{2 m_5^2+2 H m_6}{M^2} \;, \qquad  \btwo \equiv  \frac{2 H m_6}{M^2} \;, \\
  &   \bthree \equiv  \frac{4 H  m_7+2 H m_6 }{M^2} \;,  \qquad  \alpha_{\rm H} \equiv \frac{2 \left( \tilde m_4^2 - m_4^2 \right)}{M^2} \;.
\end{split}
\end{align}
For example, $\alphaT$ changes the speed of tensors to \cite{Gleyzes:2013ooa}
\be \label{alphat}
c^2_{\rm T} = 1 + \alphaT \ .
\ee
These dimensionless parameters also allow us to easily estimate the scales suppressing various interactions.  One can show that, in terms of the canonically normalized scalar field $\pi_{\rm c}$, we have for example
\be
m_3^3 \delta K \delta g^{00} \sim \frac{1}{\Lambda_3^3} (\partial \pi_{\rm c})^2 \partial^2 \pi_{\rm c} \ ,
\ee 
so that this interaction is suppressed by $\Lambda_3$, as desired.  Because $\Lambda_3$ is the scale suppressing the interactions, it is often referred to as the unitarity cutoff of the EFT.  

We would like to mention that a lot of work has been done on scalar-tensor theories in the covariant formulation, where one starts with a covariant action and then expands around the background \cite{Horndeski:1974wa, Silva:2009km, Deffayet:2011gz, Kobayashi:2011nu, Langlois:2015cwa, Crisostomi:2016czh, BenAchour:2016fzp},\footnote{For a relationship between the EFT parameters in \eqn{total_action} and a covariant formulation, see \cite{Creminelli:2018xsv}, for example. }   but in the spirit of this review, we would like to mention a few advantages of the EFT formulation.  On the theoretical side, the EFT approach is agnostic as to what kind of fundamental physics gives rise to the low-energy dynamics; the action \eqn{total_action} could indeed come from a fundamental scalar field, but it could also be the low-energy action of the longitudinal mode of a massive vector field, for example.   On the practical side, once the background is fixed as in \eqn{candlambda}, the EFT is directly an expansion in the perturbations, and so the independent free parameters are more easily identified.

%
\subsection{Linear cosmology}

Before matter starts to clump into structures, the Universe is well approximated by a homogenous expanding background with coupled linear perturbations of all of the relevant fields.  In the early Universe, this includes photons, neutrinos, baryons, and dark matter, for example.  As the Universe cools, non-relativistic matter starts to dominate the dynamics, and the dark matter starts to fall into larger and larger potential wells.  When the dark-matter overdensity reaches $\mathcal{O}(1)$, the evolution enters the non-linear regime.   

The linear regime holds through the CMB until the early times of matter domination.  This means that both CMB observables and the initial conditions for LSS can be computed with linear theory.  In this section, we discuss some modifications of CMB and linear LSS observables that result from the EFT of DE, see for example \cite{Piazza:2013pua, Hu:2013twa,  Kase:2014yya, Bellini:2014fua, Ade:2015rim, Lombriser:2015cla, Perenon:2015sla, Gleyzes:2015rua, Frusciante:2016xoj, Hu:2016zrh, Salvatelli:2016mgy, Renk:2016olm, Leung:2016xli, Pogosian:2016pwr, DAmico:2016ntq, Raveri:2014cka, Bellini:2015xja, Zumalacarregui:2016pph, Huang:2015srv}.

\begin{figure}[t]
\centerline{\includegraphics[width=0.5\textwidth]{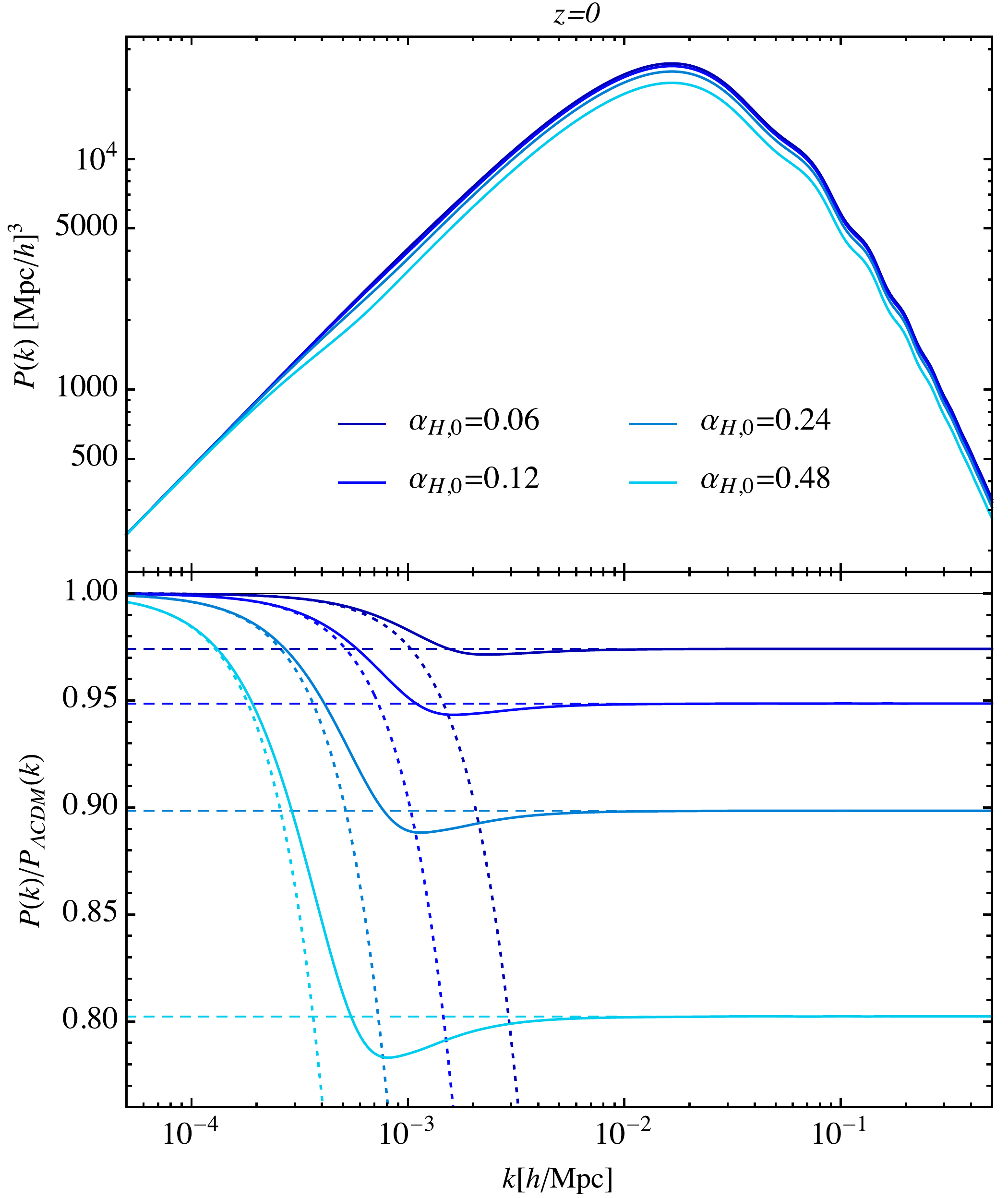}
\includegraphics[width=0.5\textwidth]{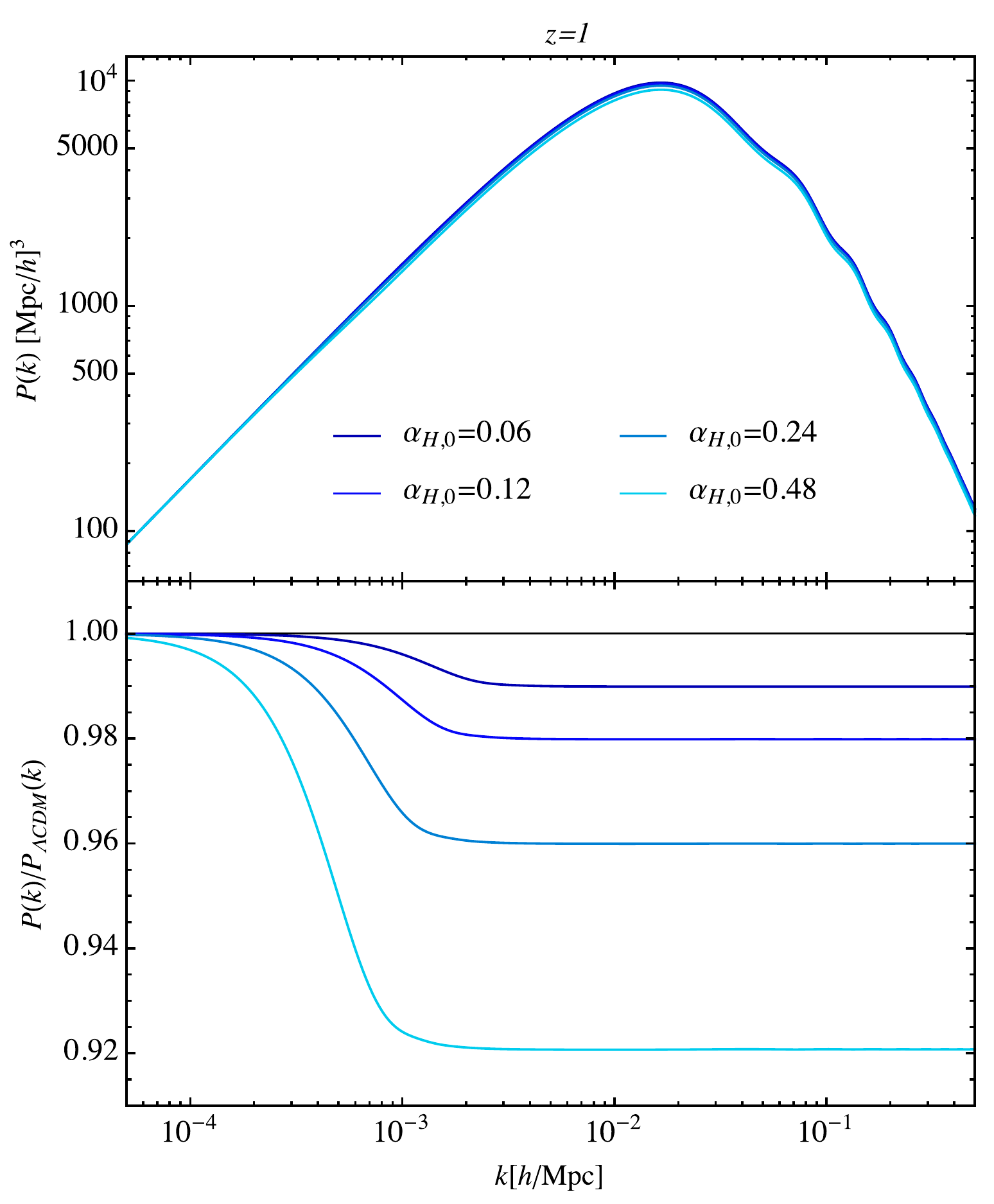}}
\caption{\footnotesize From \cite{DAmico:2016ntq}.  \emph{Original caption, slightly edited for context:}  Effect of KMM on the matter power spectrum for four different values of $\alphaH$ today, i.e.~$\alpha_{\rm H,0} = 0.06$, $0.12$, $0.24$ and $0.48$, at redshift $z=0$ (left panel) and $z=1$ (right panel). The lower plots display the ratio of these power spectra with the respective spectra for $\alphaH=0$. For comparison, the dashed and dotted lines in the left lower panel respectively show the quasi-static (dashed lines) approximation and a perturbative solution in $\alphaH$ (dotted lines).}
\label{fig:alphaH_PS}
\end{figure}

In Newtonian gauge, one can write the scalar part of the metric as 
\be \label{newtonianmetric}
ds^2 = -(1 + 2 \Phi) dt^2 + a(t)^2 (1 - 2 \Psi) d\xvec^2 \ , 
\ee
where $\Phi$ and $\Psi$ are the gravitational potentials.  Some early attempts to describe deviations from \lcdm included the phenomenological functions $\mu ( t , k)$ and $\gamma ( t , k)$, which parameterize changes in the Poisson equation
\be
-a^{-2} k^2 \Psi ( t , \kvec) = \frac{3}{2} H^2 \Omega_{\rm m} ( t ) \mu ( t , k) \delta ( t , \kvec ) \; ,
\ee
and the anisotropic stress 
\be
\gamma( t , k) = \frac{\Psi ( t , \kvec)}{\Phi ( t , \kvec)} \; ,
\ee
where $\mu= \gamma = 1 $ in $\Lambda$CDM.  The EFT of DE, however, provides a systematic way to compute relationships like these from a consistent theory in a controlled derivative expansion, specifically allowing one to see if certain parameters enter multiple observables.  For example, varying \eqn{action} with respect to $\Phi$, $\Psi$, and $\pi$ in the quasi-static limit gives the equations relevant for large-scale clustering, which has a solution like that shown in \eqn{newd2phi}.  These equations, in addition to the evolution equations for the fluid, are directly relevant for the non-relativistic evolution of the matter overdensity $\delta$.  An analogous relativistic set of equations can also be found, which is relevant for CMB anisotropies \cite{DAmico:2016ntq}.  

\begin{figure}[t]
\centerline{\includegraphics[width=0.525\textwidth]{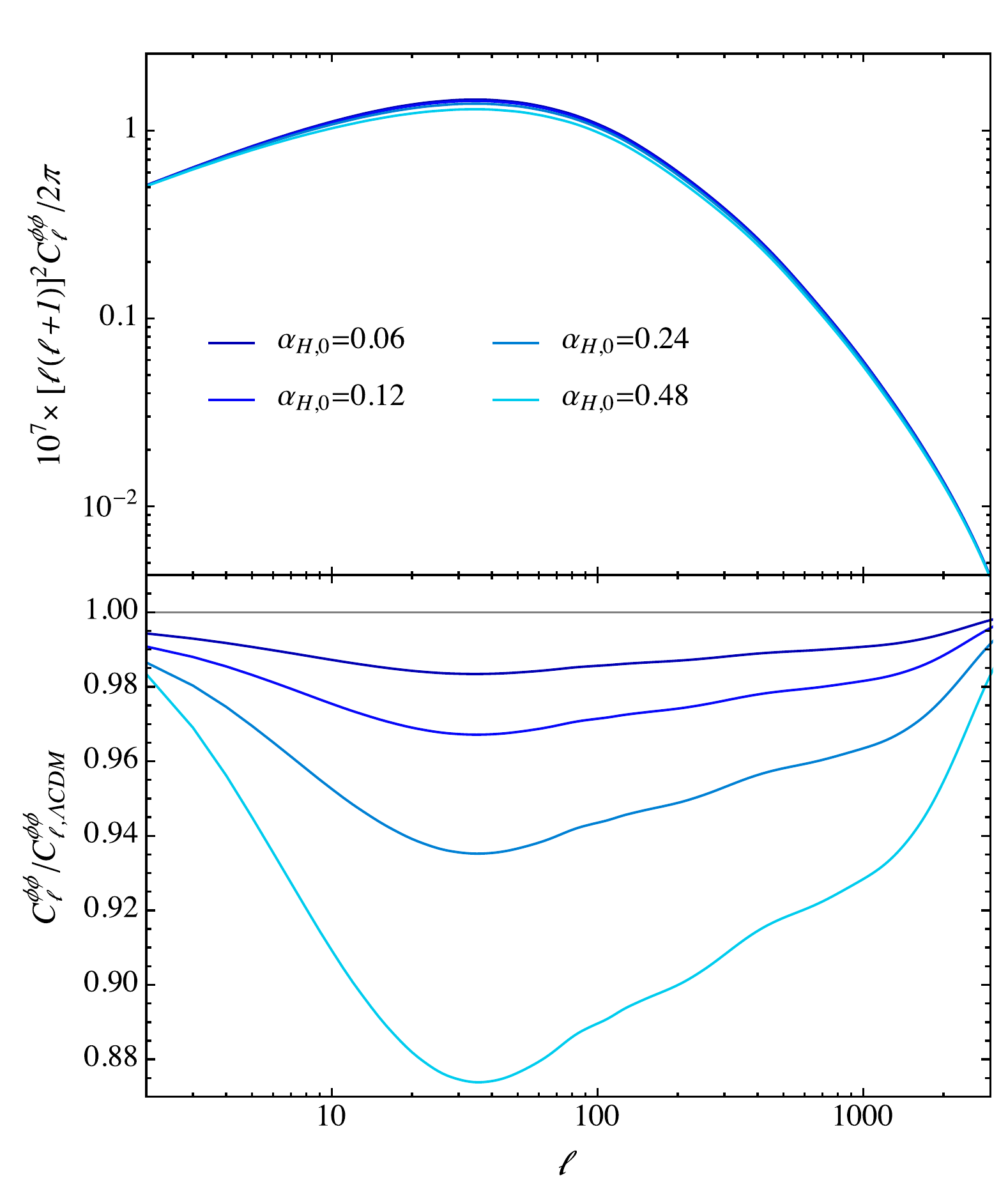}
\includegraphics[width=0.52\textwidth]{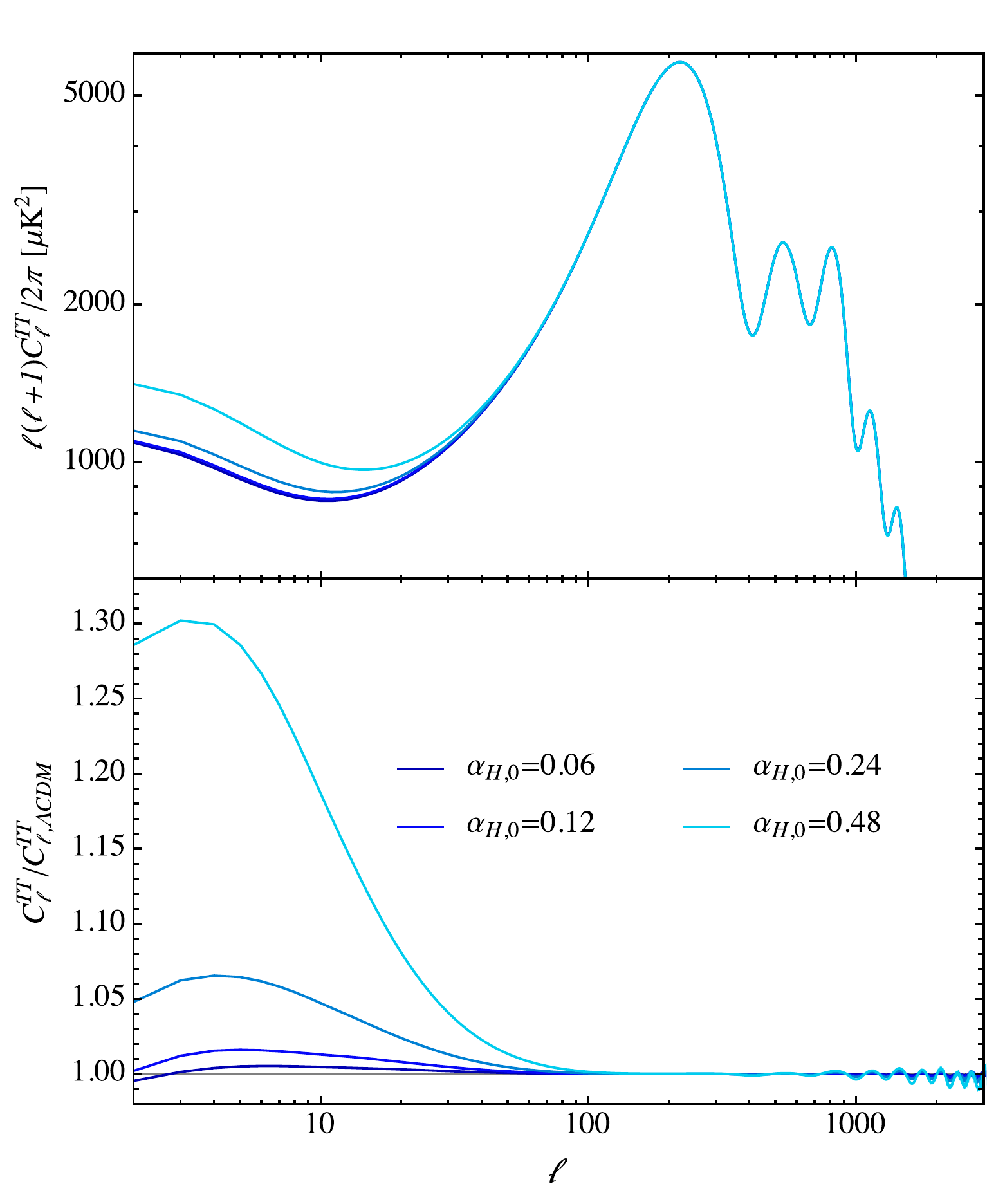}}
\caption{\footnotesize From \cite{DAmico:2016ntq}.  \emph{Original caption:} Effect of KMM ($\alphaH$) on the CMB lensing potential  (left panel) and on the CMB anisotropies (right panel) angular power spectra. The lower plots display the ratio of these angular spectra with the respective spectra for $\alphaH=0$.}
\label{fig:alphaH_CMB}
\end{figure}

As examples of DE effects, in \figref{fig:alphaH_PS} and \figref{fig:alphaH_CMB}, we show results from \cite{DAmico:2016ntq} for the modification of the matter power spectrum and the CMB, respectively, with $\alphaB = \alphaM = \alphaT = 0$, for various values of $\alpha_{\text{H},0}$, the present-day value of $\alphaH$.  The effect of $\alphaH$ is sometimes referred to as `kinetic matter mixing,' or KMM, because it results in a kinetic coupling between matter and the DE field $\pi$.  In this case, we can see that larger values of $\alphaH$ tend to suppress the matter power spectrum on small scales, suppress the CMB lensing power spectrum, and increase CMB anisotropies at small multipoles.   A number of publicly available codes have been developed to solve the linear Boltzmann equations with DE \cite{Hu:2013twa, Frusciante:2016xoj, Zumalacarregui:2016pph}.

%
\subsection{Gravitational wave propagation} \label{GWsec}
Given the explosion of measurements of GW signals coming from binary inspirals after \cite{Abbott:2016blz}, one important question is how the DE action \eqn{total_action} affects GW propagation.  As we have already seen in \eqn{alphat}, the speed of the graviton is changed by the operator $\delta \mathcal{K}_2$, proportional to $m_4^2$.   However, the observation of GW170817 and its electromagnetic counterpart GRB170817A \cite{Monitor:2017mdv} constrains the speed of gravitational waves to be the same as the speed of light to within approximately $10^{-15}$, i.e.
\be \label{speedconstraint}
|c_{\rm light} - c_{\rm GW}|  \lesssim 10^{-15} c_{\rm light} \, . 
\ee
Subsequent works used this fact to significantly constrain DE models \cite{Creminelli:2017sry, Sakstein:2017xjx, Ezquiaga:2017ekz, Baker:2017hug}.  In the language of this review, \eqn{speedconstraint} constrains the EFT of DE to have $\alpha_{\rm T} \lesssim 10^{-15}$, which for all intents and purposes means that 
\be \label{constr1}
m_4^2 = 0 \, . 
\ee
As \cite{Creminelli:2017sry} also pointed out, in order for \eqn{speedconstraint} to hold also for small changes in the cosmological background (i.e.~if the matter fraction $\Omega_{\rm m}$ were slightly different in our Universe), the EFT of DE action \eqn{total_action} must also have 
\be \label{constr2}
\tilde m_4^2 = m_5^2  \andd m_6 = \tilde m_6 = m_7 =0  \ . 
\ee
So we see that these kinds of cosmological observations can have dramatic implications for the allowed operators in the EFT of DE.

\begin{figure}[t]
\centerline{\includegraphics[width=0.28\textwidth]{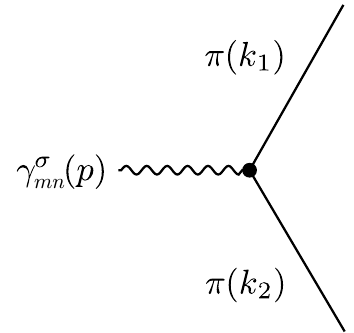}\hspace{.7in} \includegraphics[width=0.425\textwidth]{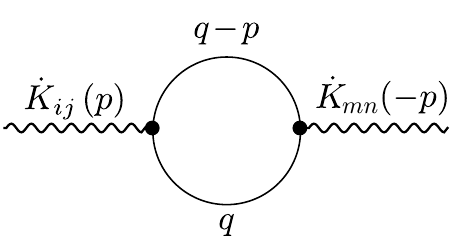}}
\caption{ \footnotesize From \cite{Creminelli:2018xsv}.  \emph{Left}: Wiggly lines represent the graviton and solid lines represent the DE field $\pi$.  When $0< c_s <1$, the graviton can decay into two scalar modes. \emph{Right}: Wiggly lines represent the graviton (through the extrinsic curvature $K_{ij}$) and solid lines represent the DE field $\pi$.  The DE field changes the graviton dispersion relation by \eqn{eq_disprel}.}
\label{decaydiag}
\end{figure}

Apart from the graviton speed, we can also use GW observations to constrain some other non-linear DE interactions.  Assuming the previously mentioned constraints \eqn{constr1} and \eqn{constr2}, the DE action \eqn{total_action} contains the term $\frac12 \tilde m_4^2(t) \, \delta g^{00}\, \left(  {}^{(3)}\!\R  +  \delta K_\mu^\nu \delta K^\mu_\nu -\delta K^2  \right)$, which leads to the interaction vertex
\begin{equation} \label{interaction}
L_{\gamma \pi \pi} =  \frac{1}{\Lambda_*^3}\ddot{\gamma}_{ij}^{\rm c} \partial_i \pi_{\rm c} \partial_j \pi^{\rm c}  \;, 
\end{equation}
between the canonically normalized graviton $\gamma_{ij}^{\rm c}$ and scalar field $\pi_{\rm c}$ \cite{Creminelli:2018xsv}, where the scale suppressing the interaction is given by 
\begin{equation}
\label{Lambdadef}
\Lambda_*^3 \equiv \MP \frac{3 m_3^6 + 4 \MP^2(c+2 m_2^4)}{2\sqrt{2}\, \tilde{m}_4^2 (\MP^2 + 2 \tilde{m}_4^2 )}  \sim \frac{\Lambda_3^3}{\alpha_{\rm H} ( 1 + \alpha_{\rm H} ) } \ .
\end{equation}
This cubic vertex allows the graviton to decay, through the diagram in \figref{decaydiag}, into two scalar modes when $0< c_s <1$, where $c_s$ is the speed of sound of the scalar.\footnote{This decay is allowed for $c_s^2 \neq 1$ because Lorentz invariance is spontaneously broken by the background evolution.}  Calling $\Gamma_{\gamma \rightarrow \pi \pi}$ the decay rate of the graviton, and demanding that gravitons do not decay over cosmological distances $\sim H^{-1}$ (since in fact we see gravitational waves), we have $\Gamma_{\gamma \rightarrow \pi \pi} H_0^{-1} \lesssim 1$ and one obtains the strong constraint \cite{Creminelli:2018xsv}
\be \label{lambdaconstraint}
\frac{\Lambda_3^3}{\Lambda_*^3} \lesssim 10^{-10} \ , 
\ee
which alternatively means 
\be
|\alpha_{\rm H}| \lesssim 10^{-10} \ ,
\ee
in GLPV theories.\footnote{For DHOST theories which have the extra parameter $\beta_1$, the constraint becomes $|\alpha_{\rm H} + 2 \beta_1 | \lesssim 10^{-10}$. }

The interaction vertex \eqn{interaction} also modifies the dispersion relation $\omega(k)$, i.e.~the dependence of the energy on wavenumber, through the Feynman diagram in \figref{decaydiag}, which is also constrained by GW measurements  \cite{Yunes:2016jcc, Abbott:2016blz}.  This interaction leads to a modification of the graviton dispersion relation \cite{Creminelli:2018xsv} 
\begin{equation}\label{eq_disprel}
\omega^2 =  k^2 - \frac{ k ^8(1-c_s^2)^2 }{480 \pi^2 \Lambda_*^6 c_s^7}\log \left( - (1-c_s^2) \frac{ k^2 }{\mu_0^{2}} - i \epsilon\right)  \ , 
\end{equation}
where $\mu_0$ is an arbitrary renormalization scale, and $\epsilon$ is a small positive parameter.  The decay rate $\Gamma_{\gamma \rightarrow \pi \pi}$ and the modified dispersion above  are indeed connected by the optical theorem 
\be
\Im \,  \omega^2 =  \Gamma_{\gamma \rightarrow \pi \pi} \omega \ ,
\ee
where the sign of $\epsilon$ is important to have the correct branch cut of the $\log$.  Current observations limit the above deviation from $\omega^2 = k^2$ to a similar level as in \eqn{lambdaconstraint} \cite{Creminelli:2018xsv}.  

While the above discussion focused on the perturbative quantum mechanical decay of the graviton, the GWs that we measure are actually classical waves with large occupation numbers, a fact that tends to further increase instabilities through Bose enhancement \cite{Creminelli:2019nok}.  As discussed in \cite{Creminelli:2019nok}, the regime of narrow resonance, where the computation is under analytic control, is able to probe $3 \times 10^{-20} \lesssim \alpha_{\rm H} \lesssim 10^{-18}$ for LIGO/Virgo type measurements, and $10^{-16} \lesssim \alpha_{\rm H} \lesssim 10^{-10}$ for a LISA type experiment.  Further study of classical effects \cite{Creminelli:2019kjy} showed that GWs can induce instabilities in the scalar field throughout the Universe unless $| \alpha_{\rm B} | \lesssim 10^{-2}$ and $|\alpha_{\rm H}| \lesssim 10^{-20}$, thus providing even stronger constraints on the EFT of DE.

Before closing this section, we should mention a few potential caveats to the above discussion.  First, since the momentum of observed GWs is close to the unitary scale $\Lambda_3$ of the EFT of DE (whose actual UV cutoff could be lower), the EFT may not actually be valid on those scales, and the above constraints would not apply \cite{deRham:2018red}.  Additionally, positivity arguments suggest that a theory with an approximate Galilean symmetry (the $\alpha_{\rm B}$ term in the EFT of DE) would have to break down at a cutoff of approximately $10^{-4} \Lambda_3$  \cite{Bellazzini:2019xts}, and so again, the above constraints would not apply.  Exceptions to this conclusion in the above discussion could be the constraints obtained in \cite{Creminelli:2019kjy}, which are relevant for GW frequencies of approximately $10^{-7 } \Lambda_3$.

%
%
\subsection{Large-scale structure}

To study effects at much smaller energies, well below the unitary cutoff $\Lambda_3$, we can turn to LSS, where the inverse length scale associated to the non-linear scale is $k_{\rm LSS} \sim (1 \text{ Mpc} )^{-1} \sim 10^{-17} \Lambda_3$, well below even the scale of current GW measurements.  Thus, LSS constraints on DE would be much more relevant if for some reason the EFT of DE breaks down near GW scales.  

In this review, we focus on DE effects in LSS in the mildly non-linear regime, and as always, we assume the Newtonian limit ($\partial^2 / H^2 \ll 1$).  To see how DE affects LSS, we start with a CDM fluid \eqn{matteraction1} coupled to DE through \eqn{totalaction}.  DE is then described by the action \eqn{total_action}, and in this section, we assume the Horndeski limit.     In this case, the form of the fluid equations is the same as in \eqn{eq:fluid}, including the stress tensor $\tau^{ij}$.

The effects of DE in \eqn{eq:fluid} come in through the modified Poisson equation, and the terms relevant for the one-loop power spectrum are (we use the scale factor $a$ as the time variable)
\begin{align} 
\begin{split} \label{newd2phi}
- \frac{ k^2}{\cH^2} \Phi (a, \kvec ) & =  \mu_{\Phi }  \frac{3 \, \omegam   }{2 } \, \delta (a, \kvec  )  + \mu_{\Phi , 2}   \left( \frac{ 3 \, \omegam}{2  } \right)^2     \int_{\kvec_1, \kvec_2}^{\kvec}  \gamma_2 ( \kvec_1 , \kvec_2 ) \delta (a, \kvec_1 ) \delta(a, \kvec_2 )  \\
&  +  \mu_{\Phi , 3}   \left( \frac{ 3 \, \omegam  }{2  } \right)^3   \int_{\kvec_1, \kvec_2, \kvec_3}^{\kvec} \gamma_3 ( \kvec_1 , \kvec_2 , \kvec_3) \delta (a, \kvec_1 ) \delta(a, \kvec_2 ) \delta (a,  \kvec_3  )  \\
& +   \mu_{\Phi , 22}   \left( \frac{ 3 \, \omegam }{2  } \right)^3    \int_{\kvec_1, \kvec_2}^{\kvec} \int_{\qvec_1, \qvec_2}^{\kvec_2}  \gamma_2 ( \kvec_1 , \kvec_2  ) \gamma_2 ( \qvec_1 , \qvec_2) \delta (a,  \kvec_1  ) \delta(a,  \qvec_1) \delta (a ,  \qvec_2  )  \ , 
\end{split}
\end{align}
where the $\mu_\Phi$ are functions of time, and are explicitly given in \cite{Cusin:2017mzw} in terms of the EFT of DE parameters in \eqn{EFTaction_masses}, and we have used the notation 
\be
\int_{\kvec_1 , \dots , \kvec_n}^{\kvec} \equiv \int \frac{ d^3 k_1}{( 2 \pi )^3} \cdots \int \frac{ d^3 k_n}{( 2 \pi )^3} \delta_D ( \kvec - \kvec_1 - \cdots - \kvec_n )  \ . 
\ee
Furthermore, the momentum dependent interaction vertices describing the effects of DE are given by\footnote{Notice that in terms of the standard mixing functions $\alpha$ and $\beta$, we have
\be
\gamma ( \kvec_1 , \kvec_2) = \frac{1}{2} \left( \alpha ( \kvec_1 , \kvec_2 ) + \alpha ( \kvec_2 , \kvec_1 ) \right) - \beta ( \kvec_1 , \kvec_2) \ . 
\ee}
\be
\label{gammas}
\begin{split}
\gamma_2 ( \kvec_1 , \kvec_2 ) & = 1 - \frac{ \big( \kvec_1 \cdot \kvec_2 \big)^2}{k_1^2 k_2^2} \ , \\
\gamma_3 ( \kvec_1 , \kvec_2 , \kvec_3 ) & = \frac{1}{k_1^2 k_2^2 k_3^2 } \Big( k_1^2 k_2^2 k_3^2  + 2 \big( \kvec_1 \cdot \kvec_2\big)  \, \big( \kvec_1 \cdot \kvec_3 \big) \, \big( \kvec_2 \cdot \kvec_3 \big)     \\
& \hspace{1in} - \big( \kvec_1 \cdot \kvec_3 \big)^2 k_2^2 -  \big( \kvec_2 \cdot \kvec_3 \big)^2  k_1^2 - \big( \kvec_1 \cdot \kvec_2 \big)^2 k_3^2 \Big) \ . 
\end{split}
\ee
For example, the second-order solution for $\delta$ (which is relevant for the one-loop power spectrum and the tree-level bispectrum, for example) has the form \cite{Bartolo:2013ws, Takushima:2013foa, Bellini:2015oua, Bellini:2015wfa, Burrage:2019afs} 
\be
\delta_{(2)}(a, \kvec  ) = \int_{\kvec_1 , \kvec_2}^{\kvec} \mathcal{F}_2 ( \kvec_1 , \kvec_2 ; a ) \delta_{(1)}(a , \kvec_1 ) \delta_{(1)} (a ,  \kvec_2 ) 
\ee
where
\be \label{f2horndeski}
\mathcal{F}_2 ( \kvec_1 , \kvec_2 ; a) = A_1(a) + A_3(a) + \frac{\hat{\kvec}_1 \cdot \hat{\kvec}_2 }{2}\left( \frac{k_1}{k_2} + \frac{k_2}{k_1} \right) + (A_2(a) - A_3(a)) \left( \hat{\kvec}_1 \cdot \hat{\kvec}_2  \right)^2
\ee
and the $A_i(a)$ are functions of time which involve integrals over Green's functions of the linear equations of motion.  As discussed in  \cite{Cusin:2017wjg}, the function $A_3$ depends on the non-linear modification $\mu_{\Phi, 2}$, and $A_1$ and $A_2$ are changed from their $\Lambda$CDM values by the linear modification $\mu_\Phi$.

\begin{figure}[t]
\centering
\includegraphics[width=13cm]{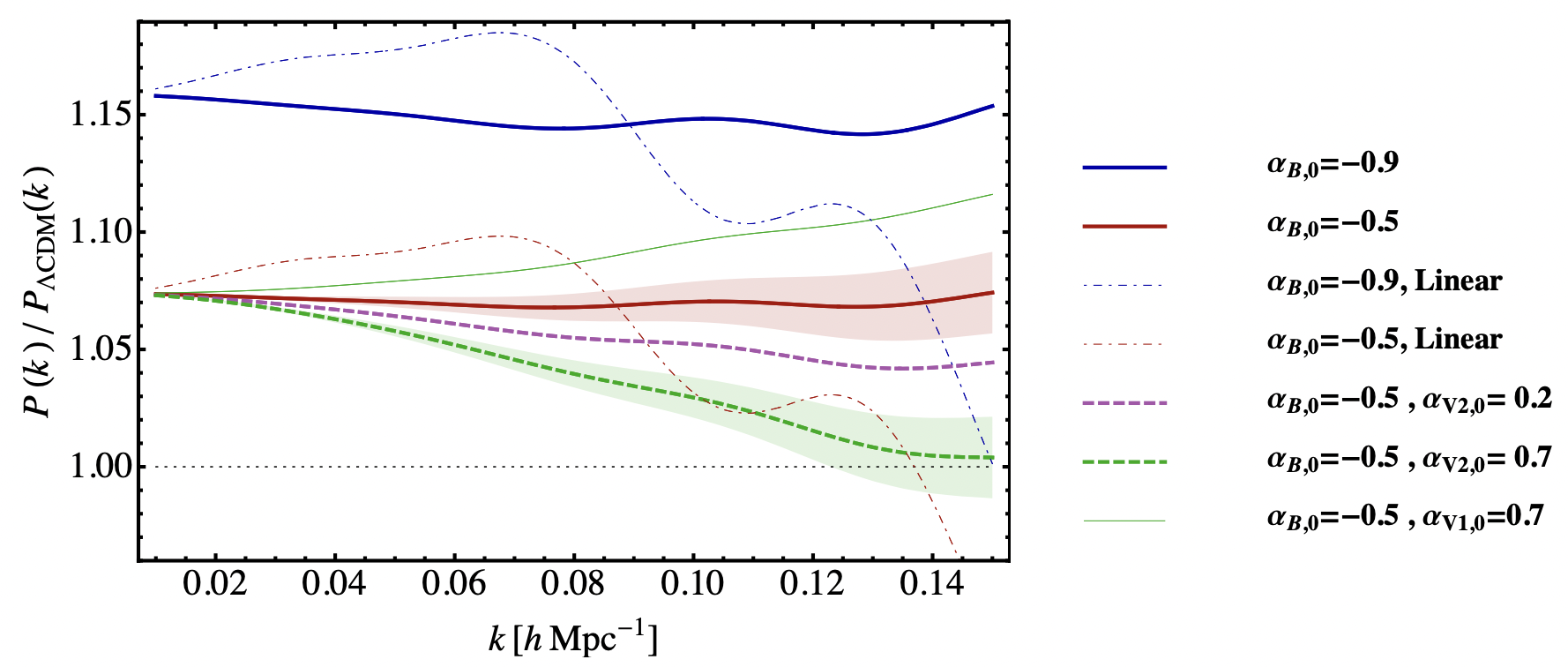} 
\caption{ \footnotesize  From \cite{Cusin:2017wjg}.  \emph{Original caption, slightly edited for brevity:}  Effect of some of the modified gravity couplings on the one-loop matter power spectrum.  The ratio between the predicted  up-to-one-loop power spectrum  with dark energy and that for $\Lambda$CDM is shown for different current values of three modified gravity parameters, $\alphaB $, $\alpha_{\rm V1}$, and $\alpha_{\rm V2} $, including LSS counterterms.  Specifically, $\alphaB $  enters at quadratic and higher order  in the action while $\alpha_{\rm V1}$ and $\alpha_{\rm V2} $ enter at cubic and higher order, so they do not modify the linear spectrum. All of the modified  gravity parameters which are not mentioned in the legend are set to zero.  Moreover, the curves labeled as ``Linear'' (thin dashed-dot lines) are the linear predictions for the corresponding values of $\alpha_{\rm B,0}$.
The bands around the dashed green and red curves are obtained by varying the amplitude of the LSS counterterm over a reasonable range.  
Since  a non-vanishing $\alphaB$ changes also the linear power spectrum with dark energy, large modifications on mildly nonlinear scales due to this parameter also imply large changes in the linear spectrum. On the other hand, $\alpha_{\rm V1}$  and $\alpha_{\rm V2}$  have a direct  effect on mildly-nonlinear scales without affecting  the linear predictions.   } \label{alphabplot}
\end{figure}

At this point, a few comments are in order.  First of all, we stress the inclusion of the stress tensor $\tau^{ij}$ from the EFT of LSS.  Since the fluid mass and momentum are conserved in Horndeski theories, the counterterms in $\tau^{ij}$ take the same form as in $\Lambda$CDM  \eqn{eq:tauij}, so that the term relevant for the computation of the one-loop power spectrum is \cite{Cusin:2017wjg}
\be
\cH^{-2} \int d^3 x\, e^{i \kvec \cdot \xvec}  \partial_i \left( \rho_{\rm m}^{-1}\partial_j \tau^{ij} (a , \xvec ) \right)  = - c_{s, \text{DE}}^2 (a) \frac{k^2}{k_{\rm NL}^2} \delta_{(1)} (a ,  \kvec ) 
\ee
where $c_{s, \text{DE}}^2(a)$ is an unknown function of time that depends on the details of short-scale clustering.  The scale $\knl$ above is the non-linear scale of structure formation, which in the presence of DE can be different from the corresponding scale in $\Lambda$CDM \cite{Cusin:2017wjg}.  The scale $\knl$ is also different from the so-called Vainshtein scale $k_V$  (discussed more in the next section), which is the scale at which the non-linear terms in the equations for the scalar field $\pi$ become important \cite{Fasiello:2017bot}.  In the limit that $k_{V} \gg \knl$, Vainshtein screening takes place on scales much smaller than gravitational non-linearities and we would expect to only see linear modifications of gravity near $\knl$.  On the other hand, if $k_V \ll \knl$, then scalar field non-linearities become important on scales much larger than those we typically use in LSS, causing the perturbative expansion assumed in this section to break down on larger scales.  Thus, the regime $\knl \lesssim k_V$ is the one in which we would expect to see new features in non-linear clustering due to DE \cite{Cusin:2017mzw}.

As mentioned in \secref{eftlsssec}, the advantage of the EFT of LSS is that one can perform precise, controllable, and improvable fits to clustering data, and as we have seen in this section, it is also the natural framework to include physics beyond the standard $\Lambda$CDM paradigm.  Work using the EFT of LSS to compare to simulations including DE was started in \cite{Bose:2018orj}.  For inclusion of a clustering quintessence DE model in the EFT of LSS, see \cite{Lewandowski:2016yce, Lewandowski:2017kes}, and for works on structure formation with DE outside of the context of the EFT of LSS, see for example \cite{Takushima:2015iha, Bose:2016qun, Bose:2018zpk, Bose:2019wuz}.

\begin{figure*}[t] 
\centering 
\hspace{-.3in} \includegraphics[width=.8\textwidth]{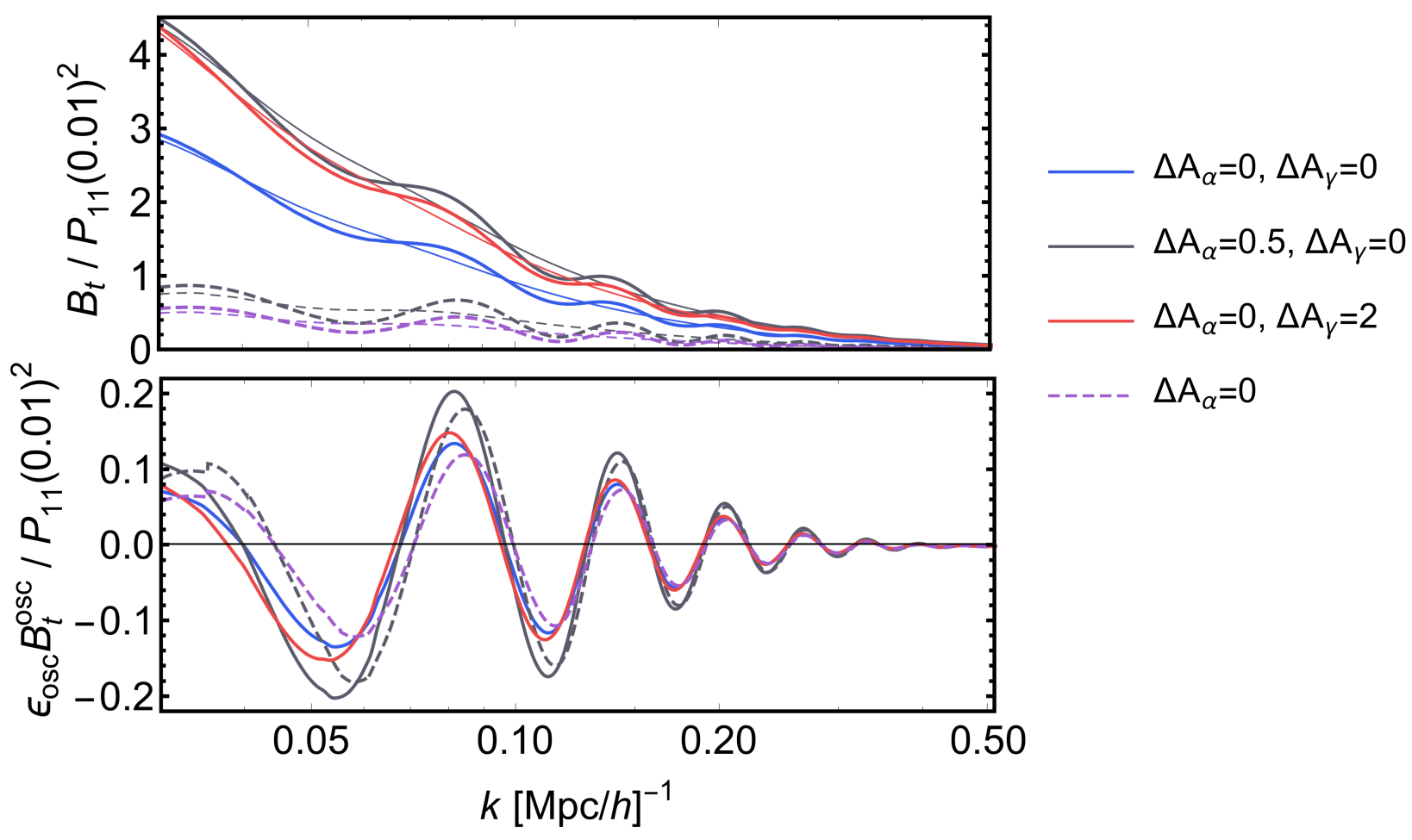} 
\caption{ \footnotesize From \cite{Lewandowski:2019txi}.  \emph{Original caption, slightly edited for brevity:} We plot various bispectra in the configuration $(\qvec , \kvec - \qvec / 2 , - \kvec - \qvec / 2)$ for $q = 0.01 \unitsk$ and $\hat{\qvec} \cdot \hat{\kvec} = 0.9$, for various values of $\Delta A_\alpha$ and $\Delta A_\gamma$.  Solid lines are the full bispectrum \eqn{fullbisp}, while dashed lines are the dominant oscillatory contributions \eqn{bispecslsubosc}.  We have also plotted the associated smooth bispectra with thin lines.  In the bottom panel, we plot the residual $\epsilonosc B_t^{\rm osc} = B_t - B_t^{\rm s}$.  In the bottom panel, we see that there are two different limiting behaviors, the purple and grey dashed lines, corresponding to $A_\alpha = 0$ and $A_\alpha = 0.5$ respectively.  The red and blue solid curves have different values of $A_\gamma$, but they have the same limiting behavior (the purple dashed line) because they have the same value of $A_\alpha$.  In all cases, the full bispectra approach the correct limiting behavior \eqn{bispecslsubosc} for $k / q \gg 1$.  We note that the large value of $\Delta A_\gamma$ needed to produce a visible difference between the red and blue curves in the plot is due to the fact that the contribution is proportional to $A_{\gamma} ( 1 - (\hat{\qvec} \cdot \hat{ \kvec} )^2)$, and we have chosen $\hat{ \qvec} \cdot \hat{ \kvec} = 0.9$ so that \eqn{bispecslsubosc} would be the dominant contribution.    } \label{bslplot}
\end{figure*}

%
\subsection{Violation of the consistency relations of LSS}

The final topic that we will mention in this section concerns the consistency relations of LSS  \cite{Peloso:2013zw, Kehagias:2013yd, Creminelli:2013mca,Valageas:2013cma,Creminelli:2013poa,Creminelli:2013nua,Kehagias:2013rpa}.  These are general, non-perturbative relations satisfied by LSS observables in various ``soft'' limits (when some momentum is taken much smaller than others).  For example, the consistency conditions are the reason that the coefficient of $\hat{\kvec}_1 \cdot \hat{\kvec}_2$ is not modified in \eqn{f2horndeski} for Horndeski theories \cite{Crisostomi:2019vhj} (a similar result holds for clustering quintessence theories \cite{Lewandowski:2017kes}).       In GLPV and DHOST theories of DE, however, the consistency relations are violated \cite{Crisostomi:2019vhj, Lewandowski:2019txi} (see also \cite{Hirano:2018uar,Hirano:2020dom}), essentially because the scalar field $\pi$ introduces a large-scale relative velocity $v^i + a^{-1} \partial_i \pi$ which cannot be removed by a coordinate transformation.    The violation of the consistency relations in DHOST theories is proportional to the relative velocity, and is computable in terms of parameters of the linear theory \cite{Lewandowski:2019txi}.  For example, the one-loop power spectrum in DHOST theories has the form 
\be \label{pir1loop}
P_{1\text{-loop}} ( k )  \Big|_{\rm IR} \sim \lambda_D \lambda_{\Delta v}^2 P_{11}(k) \int_{q \lesssim k} \frac{d^3 q}{(2 \pi)^3}\left( \frac{\qvec \cdot \kvec}{q^2} \right)^2 P_{11}(q) \ , 
\ee
where $\lambda_D$ is a parameter proportional to the DHOST parameters $\alphaH$ or $\beta_1$, and $\lambda_{ \Delta v}$ is proportional to the large-scale relative velocity \cite{Lewandowski:2019txi}.  In $\Lambda$CDM, IR dominant terms like \eqn{pir1loop} are not present in the loop expansion because of the consistency relations, but they can appear in DHOST theories.

The violation of the consistency relations in DHOST theories also affects the BAO signal \cite{Lewandowski:2019txi}.  To see how this happens, consider the tree-level bispectrum in DHOST theories
\be \label{fullbisp}
B_t ( \qvec , \kvec_1 , \kvec_2 ) = 2 F_2 ( \qvec , \kvec_1 ) P_{11} ( q ) P_{11} (  k_1 )  + \text{2 perms.} \ ,
\ee 
where
\be
F_2 ( \qvec , \kvec) = A_\alpha \alpha_s ( \qvec , \kvec) + A_\gamma \gamma ( \qvec , \kvec ) \ ,
\ee
and to describe the deviations from $\Lambda$CDM, we write
\be
A_\alpha = 1 + \Delta A_\alpha \ , \quad \text{and}  \quad A_\gamma = A_\gamma^{\Lambda \text{CDM}} + \Delta A_\gamma \ .
\ee
The fact that $A_\alpha = 1$ in $\Lambda$CDM is due to the standard consistency relations, but here we have an allowed deviation $\Delta A_\alpha$ in DHOST theories.  The value $A_\gamma^{\Lambda \text{CDM}}$, on the other hand, is cosmology dependent.  We can then consider another observable which is sensitive to IR physics, the squeezed limit of the oscillatory part of the bispectrum, $B_t^{\rm osc}$, which we define by
\be
B_t = B_t^{\rm s} + \epsilonosc B_t^{\rm osc} + \mathcal{O}(\epsilonosc^2) , 
\ee
for a small parameter $\epsilonosc \sim 0.06$.\footnote{Here, the superscripts `${\rm s}$' stands for `smooth' and `${\rm osc}$' stands for `oscillatory.'  }  The squeezed limit is given by
\be \label{bispecslsubosc}
\lim_{q \rightarrow 0} \frac{B_t^{\rm osc} ( \qvec, \kvec  - \qvec / 2, - \kvec - \qvec / 2) }{P_{11}^{\rm s} ( q ) } \approx - A_\alpha \frac{\qvec \cdot \kvec}{q^2} \frac{\qvec \cdot \kvec}{k} \frac{\partial P_{11}^{\rm osc} ( k )}{\partial k} + \mathcal{O} \left(  P^{\rm osc}_{11}(k) \right) \ ,
\ee
where $P_{11}^{\rm s}$ and $P_{11}^{\rm osc}$ are the smooth and oscillatory parts of the linear power spectrum, respectively.  Notice how, in this limit, the signal is proportional to $A_\alpha$, and not $A_\gamma$.  In $\Lambda$CDM, where $A_\alpha = 1$, this is a universal contribution, fixed by the equivalence principle \cite{Baldauf:2015xfa}.  However, in DHOST theories, we see that the size of the oscillations can change due to the deviation $\Delta A_\alpha$.  We show the various limiting behaviors in \figref{bslplot}.

%
%
\subsection{Further topics}

We close with a brief overview of some further topics in which the EFT of DE has played an important role.  One such arena is in binary mergers.  For example, the quasinormal modes in the ringdown phase can be affected by extra degrees of freedom (see e.g.~\cite{Berti:2018vdi}), and a convenient way of parametrizing these effects is through the use of EFTs.  If GR is modified by a heavy degree of freedom, then the spectrum of quasinormal modes can be described by an EFT expansion of higher derivatives of the Riemann tensor  \cite{Endlich:2017tqa}.  On the other hand, if there is an extra light degree of freedom (i.e.~dark energy), then there can be a more general deviation from GR \cite{Franciolini:2018uyq} (see also \cite{Tattersall:2017erk,Tattersall:2018nve}).   Additionally, the waveform of the inspiral phase can be modified \cite{Kuntz:2020yow}.  Possible observable effects include shifting of the quasinormal-mode spectrum, scalar radiation, and different spectra for the axial and polar modes, and the EFT approach provides a general and systematic way to search for these signatures in upcoming data.  

The EFT of DE can also be used to study Vainshtein screening and astrophysical effects.  Because GR has been so well tested in the local solar system, there is not much room in the data to modify gravity on these scales.  Thus, if we have a theory of DE that modifies gravity on large scales, some mechanism in the theory must restore GR on small scales.  One such mechanism is Vainshtein screening \cite{Vainshtein:1972sx, Babichev:2013usa}, where the non-linear couplings in the scalar field equations become large near massive sources (like the sun), and GR is restored inside of the so-called Vainshtein radius $r_V$.  However, some theories of DE can break the Vainshtein mechanism in different ways \cite{Kobayashi:2014ida,Crisostomi:2017lbg,Langlois:2017dyl,Dima:2017pwp,Crisostomi:2019yfo}, for example by having $\Phi \neq \Psi$  inside or outside of a massive object.  One can then use observational constraints, such as the Hulse-Taylor pulsar \cite{Hulse:1974eb} and the Cassini measurements \cite{Bertotti:2003rm} to constrain the parameters of the EFT of DE.   If DE changes the gravitational potentials \emph{inside} of massive objects, then astrophysical constraints related to internal star dynamics apply \cite{Sakstein:2015zoa, Crisostomi:2019yfo}. 

The final topic that we will discuss is related to \emph{theoretical} constraints on the EFT of DE.  A particularly interesting class of theoretical constraints comes from demanding positivity of the $2\rightarrow2$ scattering amplitudes (as a consequence of unitarity, analyticity, and crossing symmetry).  As mentioned in \secref{GWsec}, arguments of this kind suggest that a cosmological theory with approximate Galilean symmetry (which is generally used for Vainshtein screening) would have to break down at approximately $10^{-4} \Lambda_3$, severely limiting the applicability of the EFT on scales of cosmological interest \cite{Bellazzini:2019xts}.  Additionally, \cite{Davis:2021oce} considered more general theories with Vainshtein screening  (using positivity bounds for theories which break Lorentz invariance \cite{Grall:2021xxm}) and showed that a large number of them are inconsistent with a standard UV completion.  As another example, it was argued that EFT corrections to GR from heavy fields can only enter such that they make the graviton the fastest particle in the spectrum \cite{deRham:2019ctd}.  These kinds of theoretical constraints can shrink the allowable parameter space used to analyze data and improve bounds on EFT parameters \cite{Melville:2019wyy}, and are even stronger when including the coupling to matter \cite{deRham:2021fpu}.

\section{Conclusions}

\noindent Rapid progress in applications of EFT methods in cosmology over the last decade have shown the exciting potential that this approach has in describing cosmological perturbations in various stages in the history of the Universe, from inflation to the present day. While a lot of progress has been made so far, exploration of different EFTs remains one of the most active areas of research in theoretical cosmology and related fields. In this review we summarized the current status of the field and argued that its continued development is of paramount importance for connecting theory and observations and ultimately answering the biggest open questions in cosmology.

\bibliography{refs,short,matt_master_bib}
\bibliographystyle{utphys}

\end{document}